\documentclass[
  reprint,
  superscriptaddress,
  amsmath,
  amssymb,
  aps,
  pra,
  longbibliography
]{revtex4-2}

\usepackage{graphicx}
\usepackage{dcolumn}
\usepackage{bm}
\usepackage{hyperref}
\usepackage{xcolor}

\usepackage{newtxtext,newtxmath}

\usepackage{siunitx}
\hypersetup{
  colorlinks=true,
  linkcolor=blue,
  citecolor=blue,
  urlcolor=blue
}

\allowdisplaybreaks

\begin{document}

\title{Giant third-order polarization rotation via wave-mixing-induced symmetry breaking in a Rydberg electromagnetically induced transparency medium}

\author{Lintian Luo}
\affiliation{School of Physics, East China Normal University, Shanghai 200241, China}

\author{Yan Li}
\email{yli@phy.ecnu.edu.cn}
\affiliation{School of Physics, East China Normal University, Shanghai 200241, China}

\author{Chengjie Zhu}
\email{cjzhu@suda.edu.cn}
\affiliation{School of Physical Science and Technology, Soochow University, Suzhou 215006, China}

\author{Runbing Li}
\email{rbli@wipm.ac.cn}
\affiliation{Division of Precision Measurement Physics,
Innovation Academy for Precision Measurement Science and Technology,
Chinese Academy of Sciences, Wuhan 430071, China}
\begin{abstract}
We investigate how wave-mixing (WM)-induced symmetry breaking leads to giant third-order polarization rotation of a weak probe field in a Rydberg electromagnetically induced transparency (EIT) medium.
A far-detuned counterpropagating WM field is adiabatically eliminated and retained as an effective Raman-dressing background for the lower Zeeman manifold.
In this reduced description, the weak static magnetic field lifts the lower-state degeneracy and introduces a signed Zeeman bias between the two circularly coupled pathways, whereas the WM-induced Raman dressing reshapes the two-channel propagation symmetry that governs the subsequent Rydberg-mediated nonlinear response.
The weak-probe response is calculated using a reduced density-matrix expansion for van der Waals (vdW) correlations and self-consistent Maxwell-Bloch propagation, with the nonlinear rotation extracted by subtracting the linear propagation background.
Including WM dressing increases this nonlinear rotation from 1.06 degrees to 25.70 degrees, an enhancement of more than 24 times, for the parameters considered.
The response is nonmonotonic in WM strength and can even reverse sign, revealing that the WM field controls the propagation channels through symmetry breaking rather than merely amplifying the probe.
Eigenchannel diagnostics further show that the giant response is strongly suppressed for a single dressed-eigenchannel input and is enabled by coherent excitation of both WM-dressed propagation channels. Together with the EIT-supported Rydberg pathway and the vdW-mediated Rydberg pair-correlation response, these two-channel dynamics produce the enhanced third-order rotation.
These results identify a symmetry-breaking mechanism for Rydberg magneto-optics, with potential applications in weak-light polarimetry and all-optical polarization control.
\end{abstract}

\maketitle
\section{Introduction}
Rydberg-EIT media provide a powerful platform for weak-light nonlinear optics because long-range van der Waals (vdW) interactions between Rydberg excitations can be mapped onto optical propagation through electromagnetically induced transparency (EIT)~\cite{ref1, ref2, ref3, ref4}.
These interactions give rise to Rydberg blockade~\cite{ref5, ref6, ref7, ref8}, giant nonlocal Kerr nonlinearities~\cite{ref9, ref10, ref11}, single-photon switching and transistor operation~\cite{ref12, ref13, ref14}, quantum nonlinear optics at the few-photon level~\cite{ref15}, and neutral-atom quantum logic operations~\cite{ref16, ref17}.
For a weak probe with two circular components, polarization rotation provides a direct readout of the differential phase and attenuation accumulated by the two components.
A longitudinal magnetic field is one common way to lift the ground-state degeneracy and generate differential circular propagation, forming the basis of magneto-optical effects in atomic media~\cite{ref18, ref19, ref20, ref21}; however, an external magnetic field is not a general prerequisite for optical rotation because optical control fields can themselves create the required circular asymmetry.
Recent work has advanced balanced and integrated optical-rotation readout for atomic magnetometry~\cite{ref22, ref23}, while optical-rotation signals have also been used for vector atomic magnetometry~\cite{ref24}.
Raman-coherence-assisted nonlinear magneto-optical rotation has recently been demonstrated in cold rubidium~\cite{ref25}.

Large optically controlled rotations are already established in multilevel EIT media. Wang \textit{et al.} and Li \textit{et al.} reported rotations approaching $45^{\circ}$ with EIT-reduced probe loss~\cite{WangEITPR2006,LiEITPR2006}; hot-Na ladder EIT reached rotations up to $18^{\circ}$~\cite{HomboEIPR2012}; and V-type Rb showed EIT-associated rotation without an applied magnetic field~\cite{DasPREIT2018}. All-optical and magnetic-free control schemes provide further examples~\cite{ChoOIF2005,ZhengMFP2024}. Thus neither a large absolute rotation angle nor zero-field rotation is unique to the present system; our focus is the extracted third-order contribution and its WM-dressed Rydberg mechanism.

Wave mixing (WM) provides a complementary way to engineer optical propagation through controlled propagation asymmetry.
In atomic media, a second optical field can modify lower-state coherences and remove propagation constraints that otherwise limit the growth of a polarimetric signal~\cite{ref26, ref27}.
For comparison, accurate approximate analytical treatments have also been developed for single-probe atomic magnetometers~\cite{ref28}.
Recent theoretical work has considered polarization-qubit switching and phase control in double Rydberg-EIT media~\cite{ref29}, as well as nonlocal Rydberg enhancement of four-wave-mixing biphoton generation~\cite{ref30}.
Related experiments and theoretical modeling have examined dual-tone radio-frequency dressing of Rydberg-EIT spectra~\cite{ref31} and microwave-induced cross-Kerr nonlinearities~\cite{ref32}.
These settings are physically distinct from the far-detuned optical WM dressing considered here, but they illustrate the broader use of auxiliary fields and Rydberg interactions to engineer optical propagation and nonlinear response.
Here we use this idea in a Rydberg-EIT setting, but in a deliberately restricted form: the WM beam is far detuned and, after adiabatic elimination of the auxiliary state, is treated as a Raman dressing background that modifies the propagation symmetry of the two circular pathways.
Rydberg double-EIT and nonlinear-optical RDME treatments provide the framework for calculating the third-order response with nonlocal vdW correlations beyond a simple mean-field treatment~\cite{ref33, ref34}.
Related RDME-based studies have extended nonlocal Rydberg propagation to stable light bullets and vortices~\cite{ref35}.
Early experiments on interacting Rydberg CPT established the importance of interparticle correlations beyond simple mean-field descriptions~\cite{ref36}.
A reduced many-body density-matrix expansion was subsequently benchmarked against Rydberg CPT and EIT experiments~\cite{ref37}, while a complementary Monte Carlo description was developed for strongly interacting Rydberg-EIT media~\cite{ref38}.
This Raman dressing introduces a controlled asymmetry between the two circular propagation pathways, while the magnetic field supplies a tunable signed Zeeman bias, laying the groundwork for WM-induced symmetry breaking in the nonlinear response.

The central objective of this work is to determine how a Raman-dressed propagation background converts the Rydberg-vdW third-order response into a much larger polarization rotation.
In the present scheme, the weak magnetic field lifts the lower-state degeneracy and introduces a signed Zeeman bias, while the far-detuned WM field adds Raman-induced mixing and propagation asymmetry between the two circularly coupled pathways.
We focus on the third-order polarization rotation extracted relative to the corresponding linear propagation background, so that Raman dressing, EIT-supported Rydberg coherence, and vdW correlations can be separated while the finite-length propagation dynamics of the two circular components are retained. The far-detuned WM field modifies the two-channel propagation through Raman dressing, whereas vdW-mediated Rydberg pair correlations enter the cubic optical response. The relevant comparison is therefore the additional third-order rotation generated on the same dressed propagation background.

The main contributions are threefold.
First, the far-detuned WM beam is reduced to an effective Raman dressing field, which modifies the propagation background and breaks the symmetry between the two circular channels, thereby regulating the accumulation of the Rydberg-mediated nonlinear response.
Second, the interaction-induced rotation is extracted by comparing total and linear Maxwell-Bloch propagation on the same background, making the reported third-order rotation distinct from the absolute output rotation.
Third, controlled propagation and eigenchannel diagnostics show that, in the regime considered here, the large response is strongly suppressed for a single dressed eigenchannel and instead arises from coherent two-channel propagation.

In this work, the weak-probe response is evaluated on the WM-dressed density-matrix background, vdW-induced correlations are included with an RDME truncation, and the two circular components of the probe are propagated self-consistently using Maxwell-Bloch equations.
The quantitative manifestation of this symmetry-breaking mechanism is as follows. For the parameters studied here, the extracted third-order rotation is enhanced from \(\psi_{\rm nonlin}=+1.06^{\circ}\) in the EIT+vdW case to \(\psi_{\rm nonlin}=+25.70^{\circ}\) in the EIT+vdW+WM case, i.e., by more than a factor of 24. The value $25.70^{\circ}$ refers to this extracted third-order contribution and is not intended as a claim of a record absolute polarization rotation relative to conventional EIT schemes.
The enhancement is not monotonic in the WM strength and can turn into suppression or sign reversal outside the optimal dressing regime.
\section{Physical model and theoretical framework}
\subsection{System model and level configuration}
\begin{figure}[htbp]
\centering
\includegraphics[width=\linewidth]{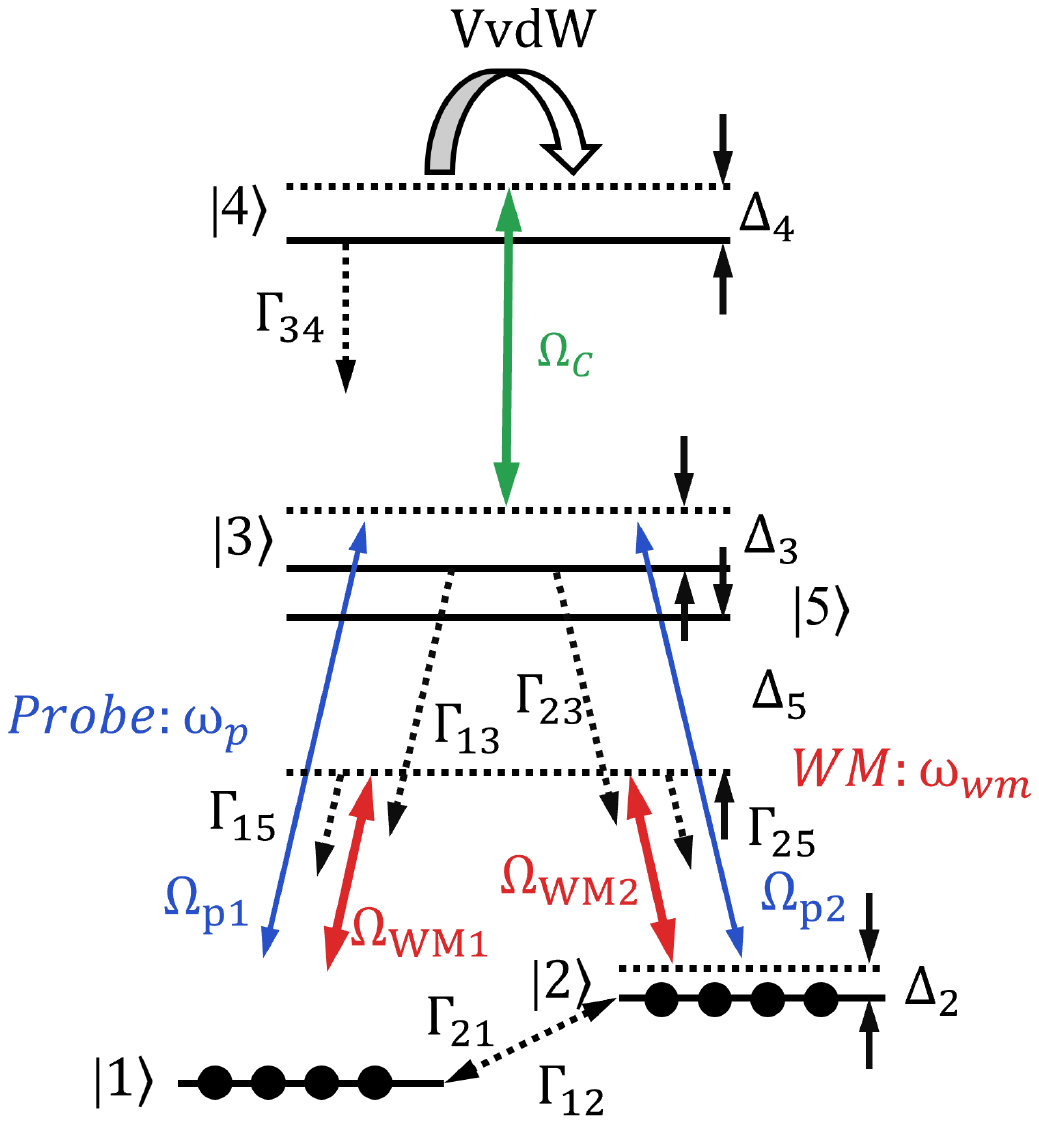}
\caption{Excitation scheme of the multilevel cold atomic system combining a wave-mixing (WM) field and long-range Rydberg interactions.
A weak magnetic field splits the two ground states $|1\rangle$ and $|2\rangle$, providing a controllable signed Zeeman bias between the two circularly coupled pathways.
The two circular components of the same probe field (blue), with Rabi frequencies $\Omega_{p1}$ and $\Omega_{p2}$, share the common carrier frequency $\omega_p$ and couple $|1\rangle$ and $|2\rangle$ to $|3\rangle$, respectively.
The control field (green), with carrier frequency $\omega_c$ and Rabi frequency $\Omega_c$, couples $|3\rangle$ to the Rydberg state $|4\rangle$ to form the dual-EIT core.
Likewise, the two circular components of the same linearly polarized far-detuned WM field (red), with Rabi frequencies $\Omega_{{\rm WM}1}$ and $\Omega_{{\rm WM}2}$, share the common carrier frequency $\omega_{\rm WM}$ and couple $|1\rangle$ and $|2\rangle$ to the auxiliary state $|5\rangle$, respectively, generating the Raman dressing used to modify the effective propagation symmetry.
The separate arrows denote polarization-selective coupling pathways rather than distinct optical frequencies.
Dashed arrows denote decay or incoherent transfer rates, and $V_{\rm vdW}$ denotes the Rydberg--Rydberg interaction.
The level scheme is schematic and not drawn to scale.}
\label{fig1}
\end{figure}

As illustrated in Fig.~\ref{fig1}, we consider an ultracold $^{85}$Rb gas ($I=5/2$) modeled by a five-level system that includes spontaneous emission.
A weak static magnetic field along the $z$ axis lifts the ground-state degeneracy and introduces a signed splitting between the two circularly coupled ground-state pathways, whose Zeeman sublevels are $|1\rangle = |5^2S_{1/2}, F = 3, m_F = 1\rangle$ and $|2\rangle = |5^2S_{1/2}, F = 3, m_F = -1\rangle$.
The intermediate and Rydberg states are $|3\rangle = |5^2P_{3/2}, F = 4, m_F = 0\rangle$ (D2 line, $\sim \SI{780}{nm}$) and $|4\rangle = |68S_{1/2}, F = 3, m_F = 1\rangle$, respectively.
An additional excited state $|5\rangle = |5^2P_{1/2}, F = 3, m_F = 0\rangle$ (D1 line, $\sim \SI{795}{nm}$) mediates the far-detuned Raman coupling between $|1\rangle$ and $|2\rangle$.
Any residual magnetic-field-induced shifts of the upper levels are absorbed into effective detunings or neglected owing to the large optical detunings.

The driving optical fields are configured as follows.

\textit{Probe field.}---A weak probe field (frequency $\omega_p$) propagates along $+z$. Its $\sigma^-$ and $\sigma^+$ polarized components drive the $|1\rangle \leftrightarrow |3\rangle$ and $|2\rangle \leftrightarrow |3\rangle$ transitions, respectively. The corresponding single-photon Rabi frequencies are $\Omega_{p1} = \mathbf{d}_{13} \cdot \hat{\bm{\epsilon}}_- \mathcal{E}_{p-} / \hbar$ and $\Omega_{p2} = \mathbf{d}_{23} \cdot \hat{\bm{\epsilon}}_+ \mathcal{E}_{p+} / \hbar$, where $\mathbf{d}_{ij}$ are the dipole transition matrix elements.

\textit{Control field.}---A strong $\sigma^+$-polarized control field (frequency $\omega_c$) counterpropagates along $-z$ and couples the $|3\rangle \leftrightarrow |4\rangle$ transition with Rabi frequency $\Omega_c$. Together, the probe and control fields establish a dual-channel Rydberg electromagnetically induced transparency (EIT) configuration~\cite{ref33, ref39, ref40}.

\textit{Wave-mixing field.}---A linearly polarized WM field (frequency $\omega_{\rm WM}$) also propagates along $-z$. Its $\sigma^-$ and $\sigma^+$ components drive the $|1\rangle \leftrightarrow |5\rangle$ and $|2\rangle \leftrightarrow |5\rangle$ transitions with Rabi frequencies $\Omega_{{\rm WM}1}$ and $\Omega_{{\rm WM}2}$, respectively. A large single-photon detuning ($\Delta_5 \gg \Delta_{2,3,4}$) suppresses direct optical absorption while establishing Raman coherence between the ground states. The Raman coupling induced by the copropagating $\sigma^{\pm}$ components of the WM beam is inherently phase-matched. For the counterpropagating probe and control fields, residual two-photon Doppler broadening is set by the probe--control wave-vector mismatch~\cite{ref41}. For $\lambda_p\simeq780~\mathrm{nm}$ and $\lambda_c\simeq480~\mathrm{nm}$, the corresponding Gaussian FWHM is $\Delta\nu_{2\gamma,D}^{\rm FWHM}\simeq18.7~\mathrm{kHz}\sqrt{T/(1~\mu\mathrm{K})}$. Comparison with the $\sim4\times10^4~\mathrm{Hz}$ characteristic WM-dressed two-photon FWHM gives a conservative crossover scale of several microkelvin.

The electric fields are expressed as
\begin{equation}
\begin{aligned}[b]
\mathbf{E}_{p} &= (\hat{\bm{\epsilon}}_+ \mathcal{E}_{p+} + \hat{\bm{\epsilon}}_- \mathcal{E}_{p-})\exp[i(k_p z - \omega_p t)] + \text{c.c.}, \\*
\mathbf{E}_{c} &= \hat{\bm{\epsilon}}_+ \mathcal{E}_{c}\exp[i(-k_c z - \omega_c t)] + \text{c.c.}, \\*
\mathbf{E}_{\rm WM} &= (\hat{\bm{\epsilon}}_+ \mathcal{E}_{{\rm WM}+} + \hat{\bm{\epsilon}}_- \mathcal{E}_{{\rm WM}-})\exp[i(-k_{\rm WM} z - \omega_{\rm WM} t)] \\*
&\quad + \text{c.c.},
\end{aligned}
\end{equation}
where $k_j = \omega_j/c$ are the wavenumbers and $\mathcal{E}$ are the slowly varying envelopes. The circular polarization unit vectors are $\hat{\bm{\epsilon}}_{\pm} = (\hat{\mathbf{x}} \mp i\hat{\mathbf{y}})/\sqrt{2}$. For the linearly polarized WM field, the Rabi frequencies of its circular components are defined as
\begin{equation}
\Omega_{{\rm WM}1} = \frac{\Omega_{\rm WM}}{\sqrt{2}}e^{-i\pi/4}, \qquad \Omega_{{\rm WM}2} = \frac{\Omega_{\rm WM}}{\sqrt{2}}e^{+i\pi/4},
\label{eq:WM_circular_decomposition}
\end{equation}
yielding equal amplitudes ($|\Omega_{{\rm WM}1}|=|\Omega_{{\rm WM}2}|=|\Omega_{\rm WM}|/\sqrt{2}$) and a relative phase of $\pi/2$.

\subsection{Hamiltonian and master equation}
Applying the electric-dipole and rotating-wave approximations (RWA)~\cite{ref33, ref42}, we set $E_1 = 0$ and $\omega_j = E_j/\hbar$.
The corresponding one- and multiphoton detunings are $\Delta_2 = -\omega_2$, $\Delta_3 = \omega_p - \omega_3$, $\Delta_4 = \omega_p + \omega_c - \omega_4$, and $\Delta_5 = \omega_{\rm WM} - \omega_5$.
Here, $\Delta_2 = 2g_F \mu_B B/\hbar$ is the Zeeman shift induced by the longitudinal magnetic field $B$, where $g_F$ is the Landé $g$-factor and $\mu_B$ is the Bohr magneton.
The three-dimensional excitation geometry and the counterpropagating $\sigma^{+}$-polarized control field together satisfy the spatial propagation constraints and angular momentum selection rules.
Defining the slowly varying transition operators $\hat{S}_{\alpha\beta} = |\beta\rangle\langle\alpha| \exp\{i[(\mathbf{k}_\beta - \mathbf{k}_\alpha)\cdot \mathbf{r} - (\omega_\beta - \omega_\alpha + \Delta_\beta - \Delta_\alpha)t]\}$, the total effective Hamiltonian for the atomic ensemble with van der Waals (vdW) interactions $V(\mathbf{r}'-\mathbf{r}) = -C_6/|\mathbf{r}'-\mathbf{r}|^6$~\cite{ref43, ref44} is given by (detailed in Appendix~A)
\begin{equation}
\begin{aligned}[b]
\hat{H}_{\text{total}} &= \mathcal{N}_a\int d^3r \bigg\{-\hbar\sum_{j=2}^{5}\Delta_j\hat{S}_{jj}(\mathbf{r},t) \\*
&\quad -\hbar\big[\Omega_{p1}^*\hat{S}_{31}(\mathbf{r},t) + \Omega_{p2}^*\hat{S}_{32}(\mathbf{r},t) + \Omega_c^*\hat{S}_{43}(\mathbf{r},t) \\*
&\quad + \Omega_{{\rm WM}1}^*\hat{S}_{51}(\mathbf{r},t) + \Omega_{{\rm WM}2}^*\hat{S}_{52}(\mathbf{r},t) + \text{H.c.}\big] \\*
&\quad + \mathcal{N}_a\int d^3r' \hat{S}_{44}(\mathbf{r}',t)\hbar V(\mathbf{r}'-\mathbf{r})\hat{S}_{44}(\mathbf{r},t)\bigg\}.
\end{aligned}
\label{eq:hamiltonian}
\end{equation}
The strong vdW repulsion suppresses multiple Rydberg excitations within a blockade radius $R_b \sim |C_6/\delta_{\text{EIT}}|^{1/6}$, where $\delta_{\text{EIT}}$ is the EIT transmission linewidth~\cite{ref11, ref45}.
The system dynamics are governed by the Lindblad master equation $\partial_t \hat{\rho} = -(i/\hbar)[\hat{H}_{\text{total}}, \hat{\rho}] + \mathcal{L}[\hat{\rho}]$.
Defining the complex decoherence rate $d_{\alpha\beta} = \Delta_\alpha - \Delta_\beta + i\gamma_{\alpha\beta}$, where $\gamma_{\alpha\beta} = (\Gamma_\alpha + \Gamma_\beta)/2 + \gamma_{\alpha\beta}^{\text{dep}}$ incorporates the spontaneous emission rate $\Gamma$ and the pure dephasing rate $\gamma^{\text{dep}}$, and applying the commutation relation $[\hat{S}_{\alpha\beta}(\mathbf{r},t), \hat{S}_{\alpha'\beta'}(\mathbf{r}',t)] = \mathcal{N}_a^{-1}\delta(\mathbf{r}-\mathbf{r}')[\delta_{\alpha\beta'}\hat{S}_{\alpha'\beta}(\mathbf{r},t) - \delta_{\alpha'\beta}\hat{S}_{\alpha\beta'}(\mathbf{r}',t)]$, we derive the single-body optical Bloch equations (OBEs) detailed in Appendix~C.

\section{Analytical method}
The calculation separates into two tasks.
First, the far-detuned WM field is reduced to an effective Raman dressing of the lower manifold.
Second, the vdW-mediated Rydberg pair-correlation response is treated with an RDME truncation of the many-body hierarchy.

\subsection{Adiabatic elimination and dimensional reduction of the wave-mixing field}
The WM field operates in the far-detuned regime and is used to add Raman-induced propagation asymmetry on top of the weak Zeeman splitting. Specifically, the single-photon detuning $\Delta_5$ relative to state $|5\rangle$ satisfies $\Delta_5 \gg \Gamma_5, |\Omega_{{\rm WM}j}|$ ($j=1,2$). With typical parameters $\Omega_{\rm WM}/2\pi = \SI{10}{MHz}$ and $\Delta_5 = 2\pi \times \SI{2000}{MHz}$, the adiabatic small parameter for the circular components is $\epsilon_W = |\Omega_{{\rm WM}j}|/|\Delta_5| \simeq 3.54\times10^{-3} \ll 1$. The probe field simultaneously operates in the weak-field regime; for $\Omega_p = 2\pi\times\SI{0.0477}{MHz}$ and $\Gamma_3 = 2\pi \times \SI{6.06}{MHz}$, the perturbation parameter is $\epsilon_p = \Omega_p/\Gamma_3 \approx 0.008 \ll 1$. These two small parameters play different roles: $\epsilon_W$ controls the adiabatic elimination of the auxiliary optical coherences involving $|5\rangle$, whereas $\epsilon_p$ orders the subsequent weak-probe response on the WM-dressed background. Given the fast transient response of $|5\rangle$, we adiabatically eliminate its density-matrix elements by setting $\partial_t \rho_{5j} = 0$~\cite{ref46, ref47}. Mixed probe--WM corrections, such as $\Omega_{p1}\rho_{53}$, scale as $\mathcal{O}(\epsilon_W \epsilon_p^2)$ and are smaller by $\epsilon_p^2$ than the retained eliminated optical coherences of order $\mathcal{O}(\epsilon_W)$. They are therefore omitted in the leading elimination. The quasi-steady-state solutions for the polarization coherences of state $|5\rangle$ are therefore
\begin{align}
\rho_{51} &\approx -\frac{\Omega_{{\rm WM}1}}{d_5}\rho_{11} - \frac{\Omega_{{\rm WM}2}}{d_5}\rho_{21}, \\
\rho_{52} &\approx -\frac{\Omega_{{\rm WM}2}}{d_5}\rho_{22} - \frac{\Omega_{{\rm WM}1}}{d_5}\rho_{12},
\end{align}
where $d_5 = \Delta_5 + i\Gamma_5/2$ (see Appendix~B for detailed derivations).

Substituting these solutions back into the dynamical equations analytically eliminates the independent degrees of freedom of the WM field. Its leading effect is a set of complex ac Stark corrections and Raman couplings in the lower manifold, on which the weak-probe response is subsequently evaluated. This dressing field introduces two key modifications: (1) ac Stark corrections: the effective detunings acquire $\Delta_{AC1} = |\Omega_{{\rm WM}1}|^2/d_5^*$ and $\Delta_{AC2} = |\Omega_{{\rm WM}2}|^2/d_5^*$; their real parts are light shifts that can be absorbed into the effective detunings or compensated by retuning the two-photon resonance, whereas their small imaginary parts represent off-resonant scattering and are retained in the complex detunings; (2) effective Raman couplings: the field induces off-diagonal coherence between the ground states, defined as
\begin{align}
\Omega_{C12} &= \frac{\Omega_{{\rm WM}1}\Omega_{{\rm WM}2}^*}{d_5^*}, \\
\Omega_{C21} &= \frac{\Omega_{{\rm WM}2}\Omega_{{\rm WM}1}^*}{d_5^*}.
\end{align}

Although the Raman couplings are generated by a second-order WM process with magnitude $\mathcal{O}(|\Omega_{\rm WM}|^2/\Delta_5)$, they are the leading finite coupling left by eliminating the far-detuned auxiliary state and are therefore retained as part of the zeroth-order dressing Hamiltonian. Treating the WM field as an additional weak perturbative variable~\cite{ref27} would enlarge the correlated vdW many-body hierarchy. We instead treat the WM field as an externally prescribed far-detuned dressing field rather than a weak perturbative variable. Consequently, the terms ``linear'' and ``third-order'' hereafter refer to the weak-probe response evaluated on this WM-dressed background.

\subsection{Perturbation expansion and RDME truncation}
Renormalizing the far-detuned WM field reduces the system to a four-level inverted-Y subspace dressed by the steady-state Raman coupling $\Omega_{C12}$.
For a weak probe field ($\epsilon_p = |\Omega_p|/\Gamma \ll 1$), we expand the single-body density matrix in a perturbation series~\cite{ref27, ref33}:
\begin{equation}
\rho_{\alpha\beta} = \rho_{\alpha\beta}^{(0)} + \rho_{\alpha\beta}^{(1)} + \rho_{\alpha\beta}^{(2)} + \rho_{\alpha\beta}^{(3)} + \mathcal{O}(\epsilon_p^4).
\end{equation}

Substituting this series into the single-body optical Bloch equations (Appendix~C) yields the following perturbation orders:
(i) Zeroth order [$\mathcal{O}(\epsilon_p^0)$]: The far-detuned WM field establishes the steady lower-manifold dressing background on which the weak-probe expansion is performed.
(ii) First order [$\mathcal{O}(\epsilon_p)$]: The off-diagonal Raman matrix elements $\Omega_{C12}$ and $\Omega_{C21}$ mix the two circular probe pathways in the first-order coherence matrix, producing the WM-dressed two-channel linear response.
(iii) Second order [$\mathcal{O}(\epsilon_p^2)$]: The probe induces population redistributions (e.g., $\rho_{33}^{(2)}$, $\rho_{44}^{(2)}$) and low-frequency coherences ($\rho_{43}^{(2)}$) via single- and two-photon interference.
(iv) Third order [$\mathcal{O}(\epsilon_p^3)$]: The third-order coherence $\rho_{4j}^{(3)}$, governing the Kerr nonlinearity, couples to the nonlocal two-body correlated integral $\mathcal{N}_a \int d^3r' V(\mathbf{r}'-\mathbf{r})\rho_{44,4j}^{(3)}(\mathbf{r}',\mathbf{r})$ through long-range vdW interactions.

This coupling introduces an unclosed BBGKY hierarchy, in which the evolution of single-body operators depends on two-body operators $\rho_{\alpha\beta,\mu\nu}(\mathbf{r}',\mathbf{r}) = \langle\hat{S}_{\beta\alpha}(\mathbf{r}')\hat{S}_{\nu\mu}(\mathbf{r})\rangle$.
According to the Heisenberg equation of motion $\partial_t\langle\hat{S}_1\hat{S}_2\rangle = \langle(i\partial_t\hat{S}_1)\hat{S}_2\rangle + \langle\hat{S}_1(i\partial_t\hat{S}_2)\rangle$, the evolution of these two-body operators inherently involves three-body operators~\cite{ref37, ref33}.
For instance, evaluating $\rho_{42,41}$ requires
\begin{equation}
\begin{aligned}[b]
i\partial_t& \langle\hat{S}_{24}(\mathbf{r}')\hat{S}_{14}(\mathbf{r})\rangle \\*
=& -D_{42}\langle\hat{S}_{24}(\mathbf{r}')\hat{S}_{14}(\mathbf{r})\rangle - \Omega_c\langle\hat{S}_{23}(\mathbf{r}')\hat{S}_{14}(\mathbf{r})\rangle \\*
& - \Omega_{C21}\langle\hat{S}_{14}(\mathbf{r}')\hat{S}_{14}(\mathbf{r})\rangle \\*
& + \mathcal{N}_a\int d^3x V(\mathbf{x}-\mathbf{r}')\langle\hat{S}_{44}(\mathbf{x})\hat{S}_{24}(\mathbf{r}')\hat{S}_{14}(\mathbf{r})\rangle \\*
& - D_{41}\langle\hat{S}_{24}(\mathbf{r}')\hat{S}_{14}(\mathbf{r})\rangle - \Omega_c\langle\hat{S}_{24}(\mathbf{r}')\hat{S}_{13}(\mathbf{r})\rangle \\*
& - \Omega_{C12}\langle\hat{S}_{24}(\mathbf{r}')\hat{S}_{24}(\mathbf{r})\rangle \\*
& + \mathcal{N}_a\int d^3x V(\mathbf{x}-\mathbf{r})\langle\hat{S}_{24}(\mathbf{r}')\hat{S}_{44}(\mathbf{x})\hat{S}_{14}(\mathbf{r})\rangle.
\end{aligned}
\end{equation}
Following the reduced many-body density-matrix expansion for interacting Rydberg gases~\cite{ref37} and its nonlinear-optical extensions~\cite{ref34, ref33}, we factorize the three-body density-matrix elements to truncate the hierarchy and obtain mathematical closure.
Recent work has continued to employ reduced-density-matrix methods in studies of nonlocal nonlinear propagation in Rydberg media~\cite{ref33, ref34, ref35, ref48}.
The factorization reads
\begin{equation}
\begin{aligned}[b]
\rho_{\alpha\beta,\mu\nu,\gamma\delta}&(\mathbf{r}'',\mathbf{r}',\mathbf{r},t) \\*
\approx& \rho_{\alpha\beta}(\mathbf{r}'',t)\rho_{\mu\nu,\gamma\delta}(\mathbf{r}',\mathbf{r},t) \\*
& + \rho_{\alpha\beta,\mu\nu}(\mathbf{r}'',\mathbf{r}',t)\rho_{\gamma\delta}(\mathbf{r},t) \\*
& + \rho_{\alpha\beta,\gamma\delta}(\mathbf{r}'',\mathbf{r},t)\rho_{\mu\nu}(\mathbf{r}',t) \\*
& - 2\rho_{\alpha\beta}(\mathbf{r}'',t)\rho_{\mu\nu}(\mathbf{r}',t)\rho_{\gamma\delta}(\mathbf{r},t).
\end{aligned}
\end{equation}
Through the RDME factorization, high-order polarizations are truncated beyond the mean-field and ground-state approximations. The second-order two-body polarizations form a closed $10 \times 10$ linear system, yielding low-order correlation elements such as $\rho_{42,41}^{(2)}$. The third-order two-body polarizations, which include the terms $\rho_{44,41}^{(3)}$ and $\rho_{44,42}^{(3)}$ that underlie the third-order propagation response, expand into a closed $16 \times 16$ complex matrix system (detailed in Appendix~D).

In solving this coupled system, the repulsive potential $V(\mathbf{r}'-\mathbf{r})$ appears on the diagonal of the evolution matrix. As the interatomic distance vanishes ($V \rightarrow \infty$), matrix inversion yields two-body terms that scale as $\rho_{44,4j}^{(3)} \propto 1/V \rightarrow 0$. This algebraic behavior reflects the Rydberg blockade mechanism, which regularizes the short-range divergence at zero interatomic separation.

\subsection{Maxwell--Bloch propagation equations and polarization dynamics}
The spatiotemporal dynamics of the dual-component ($\sigma^\pm$) probe field are governed by the macroscopic Maxwell wave equations, which encode nonlinear circular birefringence and dichroism. Under the slowly varying envelope approximation (SVEA) and plane-wave approximation, the Rabi frequency envelopes $\Omega_{p1}$ and $\Omega_{p2}$ evolve according to~\cite{ref33, ref49}
\begin{align}
i\left(\frac{\partial}{\partial z} + \frac{1}{c}\frac{\partial}{\partial t}\right)\Omega_{p1}(z,t) + \kappa_1 \rho_{31}(z,t) &= 0, \\
i\left(\frac{\partial}{\partial z} + \frac{1}{c}\frac{\partial}{\partial t}\right)\Omega_{p2}(z,t) + \kappa_2 \rho_{32}(z,t) &= 0,
\end{align}
where $\kappa_j = \mathcal{N}_a|d_{j3}|^2\omega_p/(2\epsilon_0 c\hbar)$ is the coupling constant, $\mathcal{N}_a$ is the atomic density, and $d_{j3}$ is the relevant dipole matrix element.

In the continuous-wave (CW) steady state ($\partial_t = 0$), the truncated polarization relative to the WM-dressed background is $\rho_{3j} \simeq \rho_{3j}^{(1)} + \rho_{3j}^{(3)}$~\cite{ref33, ref49}. The first-order term $\rho_{3j}^{(1)}$ incorporates the WM-induced Raman dressing $\Omega_{C12}$, establishing an asymmetric background for linear dispersion and absorption. The third-order term $\rho_{3j}^{(3)}$ contains the vdW-mediated Rydberg pair-correlation integral and provides the third-order propagation contribution.

To quantify the dual-polarization evolution, we define the four Stokes parameters of the output field~\cite{ref27, ref50, ref51}:
\begin{align}
S_0(z) &= |\Omega_{p1}|^2 + |\Omega_{p2}|^2, \\
S_1(z) &= 2\text{Re}(\Omega_{p1}^*\Omega_{p2}), \\
S_2(z) &= 2\text{Im}(\Omega_{p1}^*\Omega_{p2}), \\
S_3(z) &= |\Omega_{p1}|^2 - |\Omega_{p2}|^2.
\end{align}
The rotation angle used below is
\begin{equation}
\psi(z)=\frac{1}{2}\operatorname{atan2}\!\left(S_2(z),S_1(z)\right),
\end{equation}
which is equivalent to \(\psi=\frac{1}{2}[\arg(\Omega_{p2})-\arg(\Omega_{p1})]\) with a continuous phase branch assigned along the propagation direction.
The resulting optical rotation can be accessed experimentally using balanced or integrated polarimetry~\cite{ref22, ref23}.

All rotations compared below are evaluated with the same input fields, propagation length, and continuous phase branch.
The total rotation contains both the linear propagation background and the third-order weak-probe contribution, whereas the nonlinear value quoted in the results denotes the background-subtracted third-order contribution.
Thus the reported enhancement refers to the extracted nonlinear rotation, not to the absolute output rotation alone.

While the constant-amplitude approximation is often employed to simplify spatial integration, the WM-induced circular-component asymmetry can couple nonlinear phase modulation with nonlinear absorption [through $\text{Im}(\chi^{(3)})$]. To accurately capture these attenuation effects, we substitute the self-consistently calculated steady-state density matrix into the coupled propagation equations and perform numerical integration along the $z$ axis using the fourth-order Runge-Kutta (RK4) method, rather than relying on static analytical phase accumulation.

\section{Results}
We numerically simulate weak-probe propagation in an ultracold rubidium atomic gas using the microscopic many-body model described above. Unless otherwise stated, detunings, Rabi frequencies, and decay or dephasing rates are angular frequencies; values are quoted through their $2\pi$-scaled frequencies in MHz. The physical parameters used in the numerical calculation are as follows: medium length $L = 15\,\mathrm{mm}$; atomic density $\mathcal{N}_a = 8 \times 10^{16}\,\mathrm{m}^{-3}$; ground-state splitting $\Delta_2 = 2\pi \times \SI{0.0042}{MHz}$; Rydberg control field $\Omega_c = 2\pi \times \SI{6.5}{MHz}$ with detuning $\Delta_4 = 2\pi \times \SI{0.18}{MHz}$; dual-EIT probe detuning $\Delta_3 = 2\pi \times \SI{100}{MHz}$; and dispersion strength $C_6 = -2\pi \times 625.6\,\text{GHz}\cdot\mu\text{m}^6$~\cite{ref43, ref44}. The linearly polarized WM field operates at $\Omega_{\rm WM} = 2\pi \times \SI{10}{MHz}$ with detuning $\Delta_5 = 2\pi \times \SI{2000}{MHz}$; its two circular components, given by Eq.~\eqref{eq:WM_circular_decomposition}, each have magnitude $2\pi\times(10/\sqrt{2})\,\mathrm{MHz}$. The initial probe amplitudes are $\Omega_{p1}(0) = \Omega_{p2}(0) = 2\pi\times\SI{0.0477}{MHz}$, corresponding to $\Omega_p/\Gamma_3\simeq0.008$. The D-line excited-state decay rates are $\Gamma_3 = 2\pi \times 6.06\ \mathrm{MHz}$ and $\Gamma_5 = 2\pi \times 6.06\ \mathrm{MHz}$~\cite{ref52, ref53}. The remaining decay and dephasing parameters are chosen as $\Gamma_4 = 2\pi \times 0.02\ \mathrm{MHz}$ and $\Gamma_{12}=\Gamma_{21} = 2\pi \times 0.0016\ \mathrm{MHz}$. The branching ratios are $\Gamma_{13} = \Gamma_3/2$, $\Gamma_{34} = \Gamma_4$, and $\Gamma_{15} = \Gamma_5/2$.

These parameters define a representative rather than globally optimized operating point. The $68S$ state lies within an experimentally demonstrated high-$n$ nonlinear-optical regime in cold $^{85}$Rb; Sinclair \textit{et al.} reported Rydberg-mediated cross-Kerr measurements including $68S$ at densities of the same order as used here~\cite{ref10}. A closely related cold-$^{85}$Rb RDME study also used $68S$ with the same nominal vdW coefficient $C_6=-2\pi\times625.6\,\mathrm{GHz}\,\mu\mathrm{m}^6$~\cite{ref33}. The weak-probe and far-detuned-WM conditions, $\Omega_p/\Gamma_3\simeq0.008\ll1$ and $|\Omega_{{\rm WM}j}|/|\Delta_5|\simeq3.54\times10^{-3}\ll1$, are well satisfied. The nominal $\Delta_2/2\pi=4.2~\mathrm{kHz}$ and $\mathcal{N}_a=8\times10^{16}\,\mathrm{m}^{-3}$ are therefore treated as representative reference points and are scanned explicitly in Fig.~\ref{fig6}(a) and (b).

The interaction length $L=15\,\mathrm{mm}$ is compatible with cold-atom platforms: elongated MOTs with longitudinal extents up to $26\,\mathrm{mm}$ have been realized~\cite{ref54}, and cold-atom memories have reached $\sigma_z=9\,\mathrm{mm}$~\cite{ref55}. Thus, $15\,\mathrm{mm}$ provides a realistic length for nonlinear propagation and accumulation.

The rotation angles calculated under two representative propagation configurations are summarized in Table~\ref{tab1}:
\begin{table}[htbp]
\caption{Calculated rotation angles for the EIT+vdW and EIT+vdW+WM configurations.
Here \(\psi_{\rm total}\) is the propagated total rotation, \(\psi_{\rm lin}\) is the corresponding linear background rotation, and \(\psi_{\rm nonlin}\) is the extracted third-order rotation.}
\label{tab1}
\begin{ruledtabular}
\begin{tabular}{lccc}
Configuration & Total & Linear & Third-order \\
& $\psi_{\text{total}}$ & $\psi_{\text{lin}}$ & $\psi_{\rm nonlin}$ \\
\hline
EIT+vdW & $-3.38^\circ$ & $-4.44^\circ$ & $+1.06^\circ$ \\
EIT+vdW+WM & $-14.11^\circ$ & $-39.81^\circ$ & $+25.70^\circ$ \\
\end{tabular}
\end{ruledtabular}
\end{table}

\subsection{WM-dressed propagation and probe power balance}
The linear background rotation in Table~\ref{tab1} includes the WM-induced Raman cross-coupling in the dressing-field convention.
The third-order rotation is obtained as $\psi_{\rm nonlin}=\psi_{\rm total}-\psi_{\rm lin}$ by subtracting the linear background from the total rotation.
This definition isolates the interaction-induced part of the probe rotation on the selected propagation background.

Without WM dressing, the EIT+vdW configuration remains close to a symmetric dual-path response.
The zeroth-order ground-state coherence satisfies $\rho_{21}^{(0)}=0$, and the vdW-induced rotation accumulates only weakly over the $15\,\mathrm{mm}$ medium.
With the far-detuned WM field, Raman dressing produces a steady lower-manifold coherence and reshapes the relative phase and amplitude evolution of the two circular probe components.

\begin{figure*}[htbp]
\centering
\begin{minipage}{0.48\textwidth}
\centering
\includegraphics[width=\textwidth]{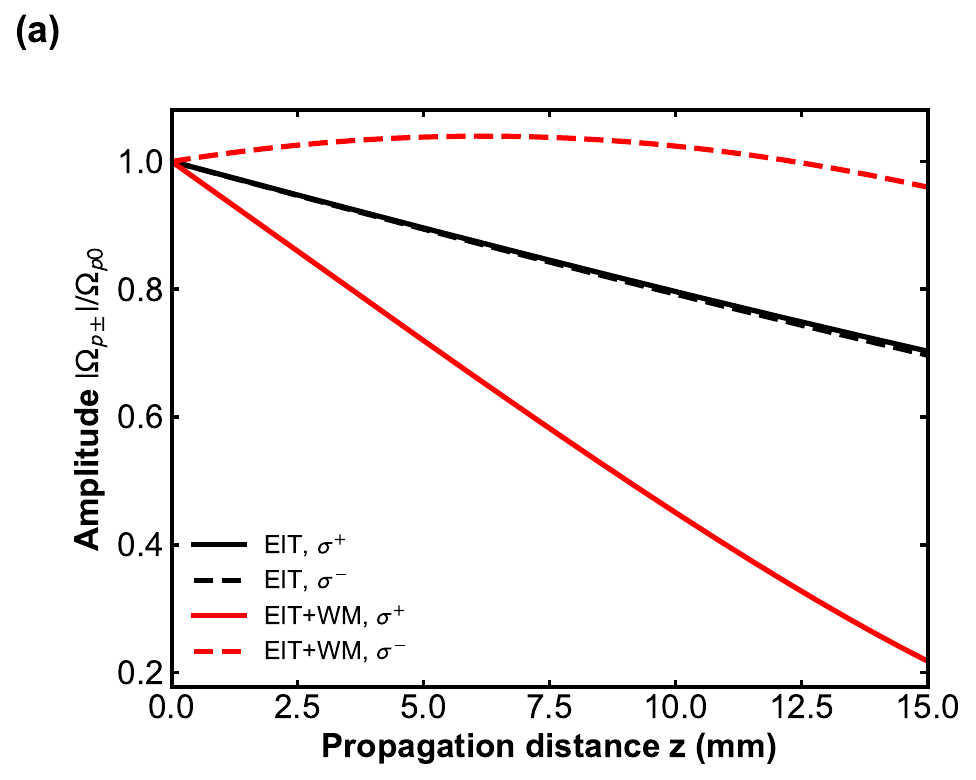}\\
\end{minipage}
\hfill
\begin{minipage}{0.48\textwidth}
\centering
\includegraphics[width=\textwidth]{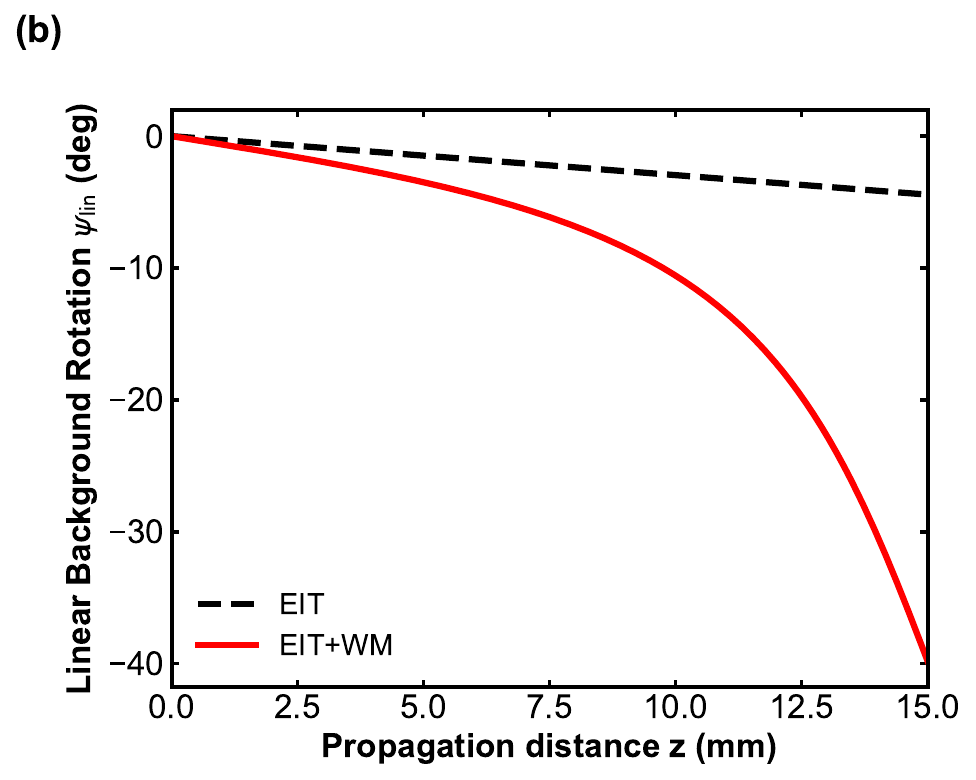}\\
\end{minipage}
\\ \vspace{0.3cm}
\begin{minipage}{0.48\textwidth}
\centering
\includegraphics[width=\textwidth]{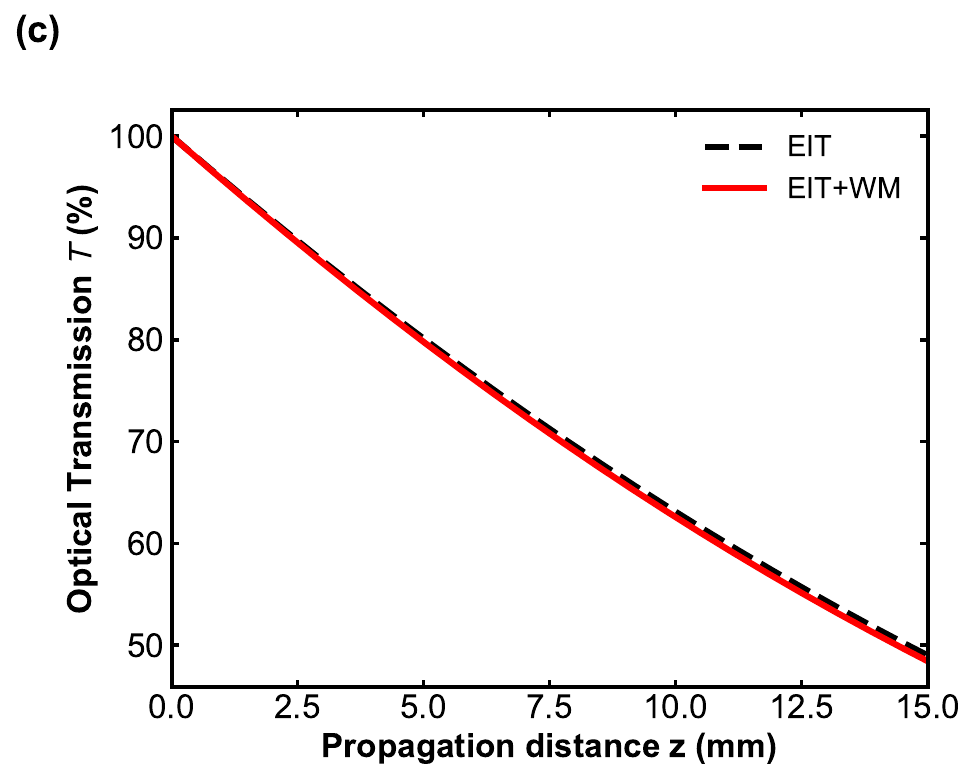}\\
\end{minipage}
\hfill
\begin{minipage}{0.48\textwidth}
\centering
\includegraphics[width=\textwidth]{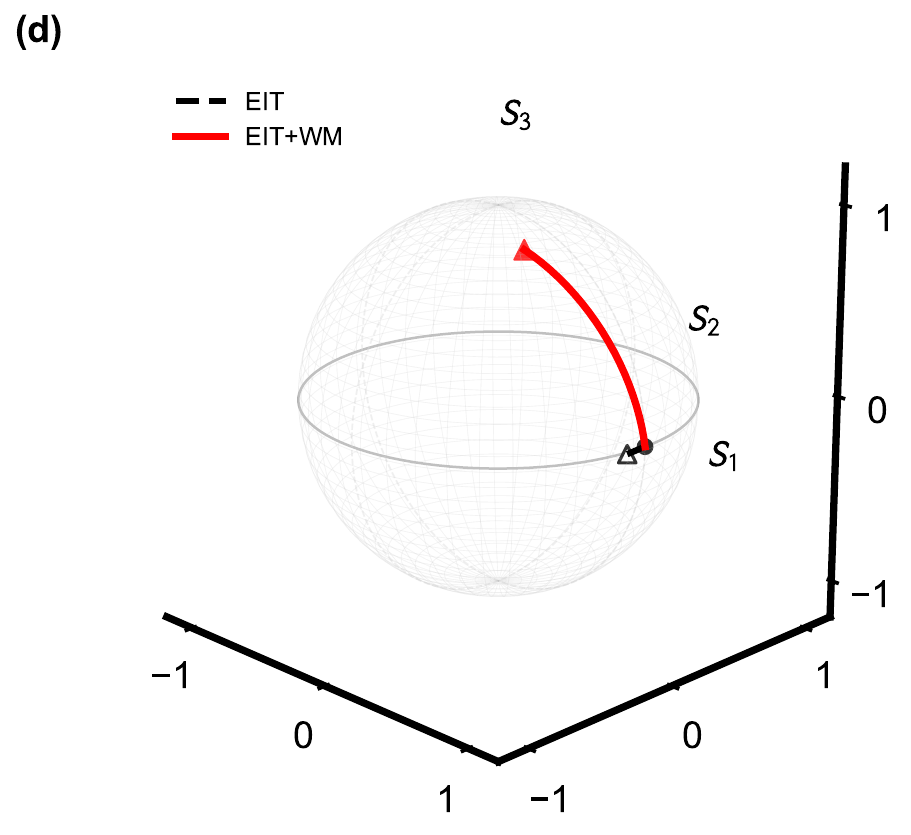}\\
\end{minipage}
\caption{WM-dressed polarization propagation for \(L=15\,\mathrm{mm}\).
Panels (a), (c), and (d) use the self-consistent total propagation including the third-order vdW response, whereas panel (b) shows the extracted first-order background.
(a) Normalized electric-field amplitudes of the two circular components; black and red curves denote EIT and EIT+WM, while solid and dashed curves denote \(\sigma^+\) and \(\sigma^-\), respectively.
(b) Linear background rotation \(\psi_{\rm lin}\).
(c) Total probe transmission obtained from the sum of the two circular-component intensities.
(d) Poincar\'{e}-sphere trajectories; triangles mark the output states at \(z=15\,\mathrm{mm}\).}
\label{fig2}
\end{figure*}

Figure~\ref{fig2}(a) shows that, in the total propagation calculation, the two circular components remain nearly balanced in the EIT+vdW case but follow distinct attenuation paths when the WM field is applied.
The extracted first-order background rotation reaches $-39.81^\circ$ at the exit [Fig.~\ref{fig2}(b)], showing that the far-detuned WM field strongly reshapes the circular-component propagation background through Raman dressing.
At the same time, Fig.~\ref{fig2}(c) shows that adding the WM field does not increase the total transmitted probe intensity compared with the EIT baseline.
Thus the WM field primarily modifies the relative phase and amplitude evolution of the two circular components through the far-detuned dressing pathway.
The appreciable attenuation in Fig.~\ref{fig2}(c) has four main contributions: residual differential absorption because the two circular components share one carrier frequency while the lower states are split by $\Delta_2$~\cite{ref39, ref41}; finite optical linewidths and ground-state dephasing; the small dissipative correction retained after eliminating the far-detuned auxiliary state; and nonlinear absorption from the complex Rydberg-mediated third-order polarization~\cite{ref49}. All are included self-consistently in the coupled Maxwell propagation. The attenuation is therefore an intrinsic part of the coupled dispersive--dissipative propagation, not an additional phenomenological loss.
The Poincar\'{e}-sphere trajectory in Fig.~\ref{fig2}(d), obtained from the same total propagation calculation, shows the resulting evolution of the full polarization state after the third-order vdW response is included.
Together, Fig.~\ref{fig2} establishes the WM-dressed propagation background from which the third-order contribution is extracted.
This is a background-control step; the nonlinear enhancement is quantified next from the extracted rotation.

\subsection{Giant third-order rotation in the WM-dressed medium}
Having established that the WM field reshapes the polarization propagation without supplying appreciable net probe power, we next isolate the third-order part of the rotation.
For the parameters used here, the WM-dressed background permits a much larger accumulation of the vdW-induced third-order response.
Table~\ref{tab1} shows that \(\psi_{\rm nonlin}\) increases from \(+1.06^\circ\) in the EIT+vdW case to \(+25.70^\circ\) in the EIT+vdW+WM case.
This nonlinear rotation is obtained by subtracting the corresponding WM-dressed linear background from the total propagated rotation.

\begin{figure}[htbp]
\centering
\includegraphics[width=\linewidth]{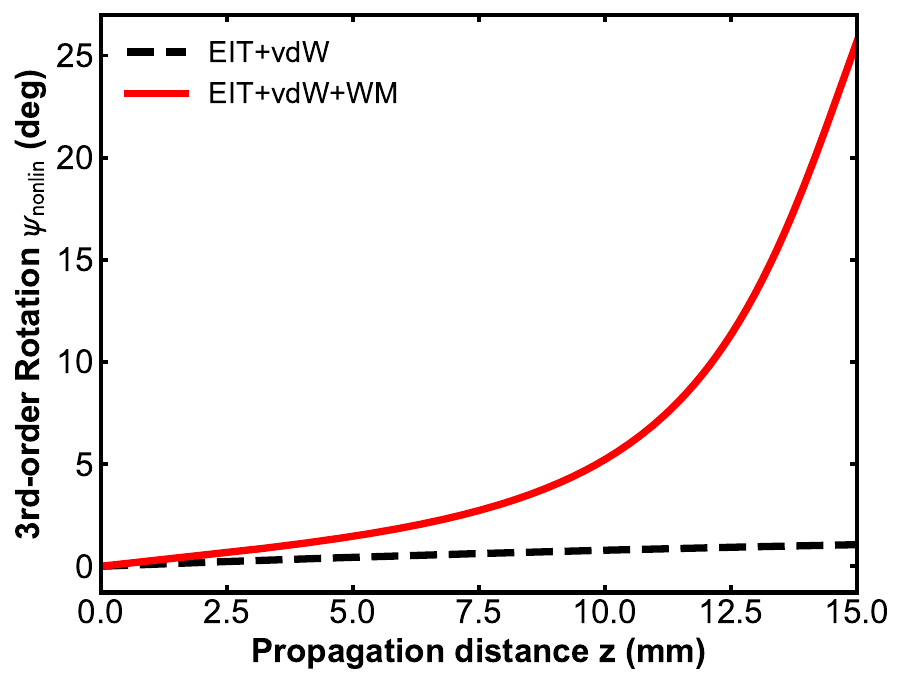}
\caption{Spatial accumulation of the extracted third-order rotation, \(\psi_{\rm nonlin}=\psi_{\rm total}-\psi_{\rm lin}\).
The EIT+vdW case gives only a weak accumulated rotation, whereas the EIT+vdW+WM case reaches \(+25.70^\circ\) at \(z=15\,\mathrm{mm}\).}
\label{fig3}
\end{figure}

Figure~\ref{fig3} is the main benchmark for the extracted nonlinear rotation.
The EIT+vdW curve accumulates only weakly, whereas the WM-dressed curve grows rapidly after Raman-induced propagation asymmetry develops.
The relevant change is the background-subtracted third-order accumulation, not merely a change in the total output rotation.
Although one circular component can be strongly attenuated, the normalized-amplitude analysis in Fig.~\ref{fig2}(a) remains relevant for coherent detection.
For the more strongly attenuated circular component, the output normalized field amplitude is about \(0.2\), corresponding to about \(4\%\) of that component's input intensity, while the total transmitted probe power in panel (c) remains much larger.
Such residual coherent fields can in principle be detected using balanced or integrated polarimetry~\cite{ref22, ref23}.
Quantum-enhanced small-angle protocols~\cite{ref56} and homodyne receivers~\cite{ref57} provide alternative sensitivity strategies.
A quantitative sensitivity prediction would additionally require the photon flux, interference visibility, detector efficiency, measurement bandwidth, and technical-noise spectrum.
The calculation is performed in the weak-probe regime, with \(\Omega_p/\Gamma_3\simeq0.008\ll1\). The low-power scaling test presented below is consistent with a cubic contribution over the experimentally relevant weak-power interval, without constituting a formal all-orders convergence proof.

\subsection{Mechanism analysis: Roles of the WM field, EIT channel, and vdW interaction}
\label{sec:mechanism}
\begin{figure*}[htbp]
\centering
\begin{minipage}{0.48\textwidth}
\centering
\includegraphics[width=\textwidth]{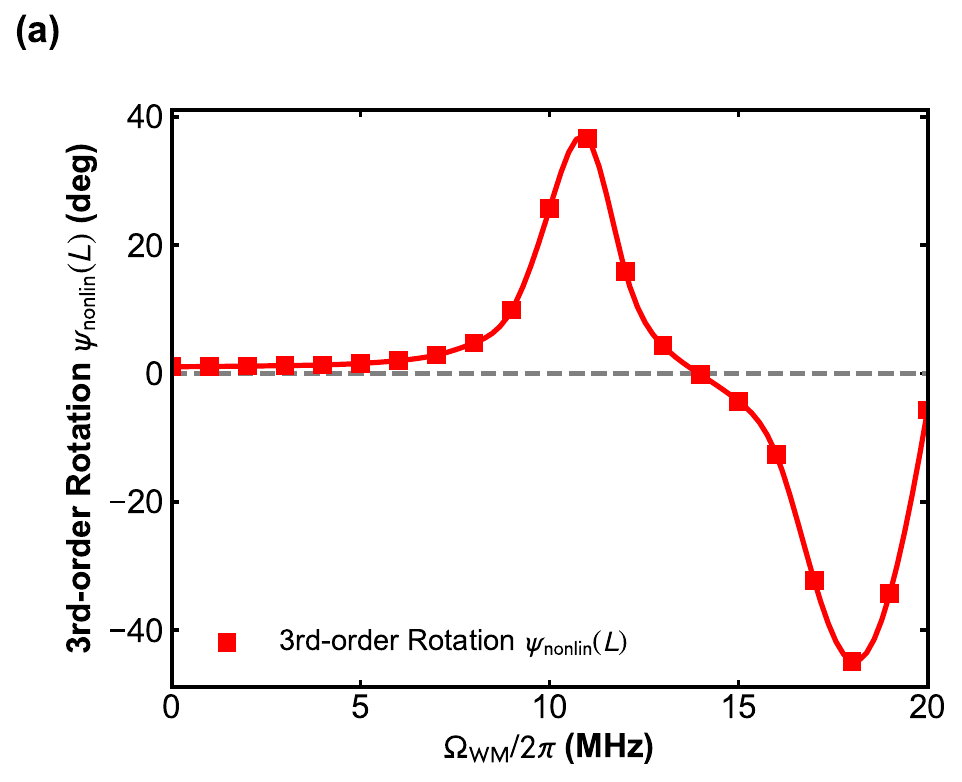}
\end{minipage}
\hfill
\begin{minipage}{0.48\textwidth}
\centering
\includegraphics[width=\textwidth]{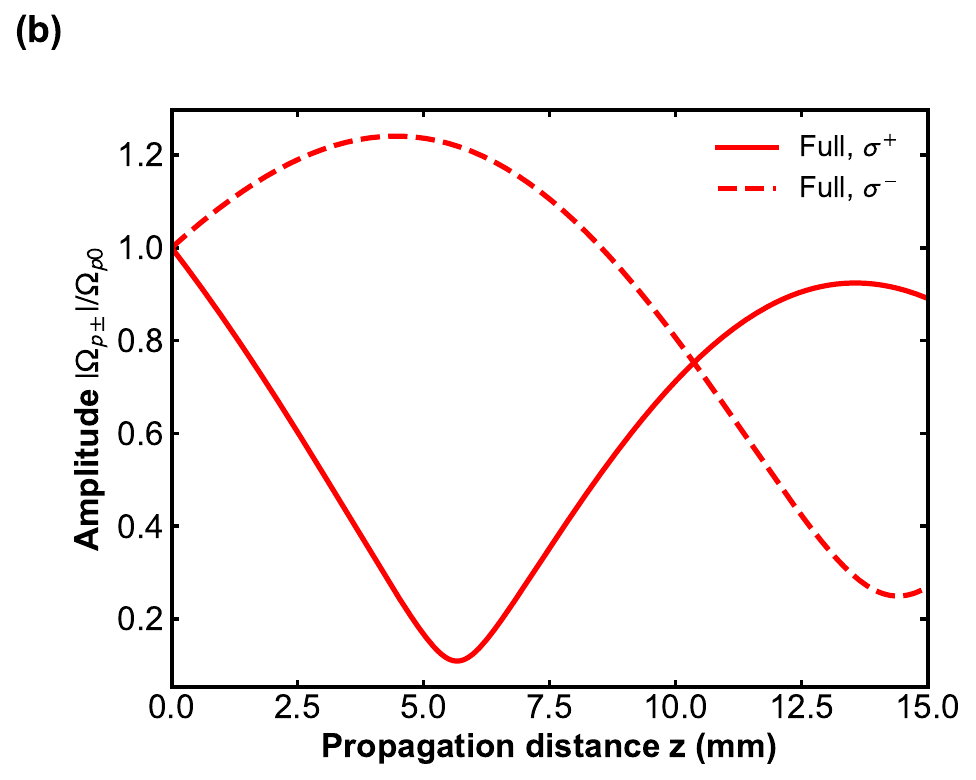}
\end{minipage}
\\[0.30cm]
\begin{minipage}{0.48\textwidth}
\centering
\includegraphics[width=\textwidth]{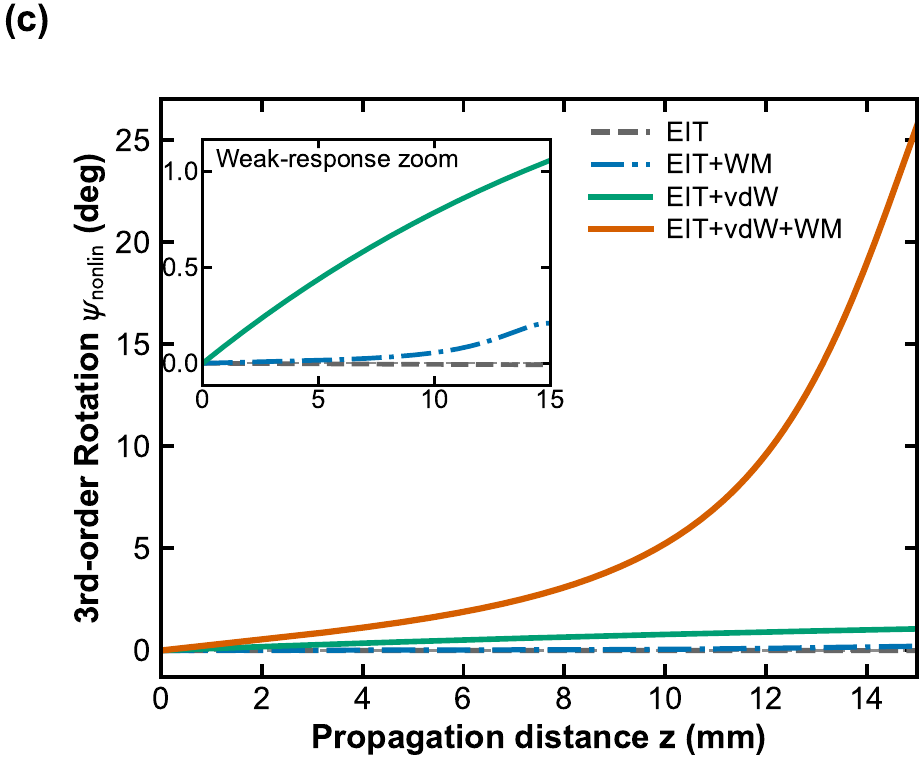}
\end{minipage}
\hfill
\begin{minipage}{0.48\textwidth}
\centering
\includegraphics[width=\textwidth]{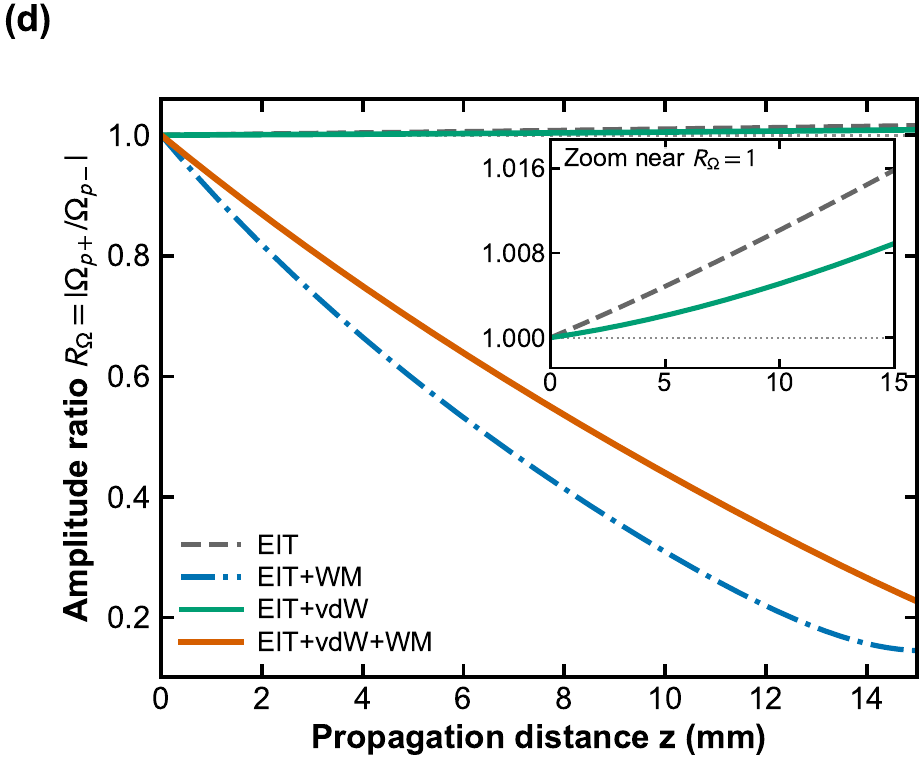}
\end{minipage}
\caption{Mechanism diagnostics for WM dressing, EIT coupling, and vdW interactions.
(a) Output third-order rotation versus WM Rabi frequency, showing a nonmonotonic response and sign reversal under strong dressing.
(b) Normalized circular-component amplitudes at \(\Omega_{\rm WM}/2\pi=20\,\mathrm{MHz}\), where strong WM dressing produces two-channel beating.
(c) Third-order rotation for EIT, EIT+WM, EIT+vdW, and EIT+vdW+WM.
(d) Corresponding amplitude ratio \(R_\Omega(z)=|\Omega_{p+}(z)/\Omega_{p-}(z)|\).}
\label{fig4}
\end{figure*}

The roles of the WM field, the EIT channel, and the vdW interaction can be separated by controlled propagation calculations, as summarized in Fig.~\ref{fig4}.

Figure~\ref{fig4}(a) shows the output third-order rotation as a function of \(\Omega_{\rm WM}/2\pi\) for the full EIT+vdW+WM configuration.
The response first grows, then decreases and changes sign under stronger dressing.
This nonmonotonic behavior is important: the WM field does not simply increase the nonlinear susceptibility or supply an additional nonlinear source.
Instead, it changes the propagation background on which the vdW-induced third-order source accumulates.

The strong-dressing example in Fig.~\ref{fig4}(b) illustrates why the response can decrease at large WM strength.
At \(\Omega_{\rm WM}/2\pi=20\,\mathrm{MHz}\), the two circular components undergo oscillatory exchange during propagation.
This beating is the field-level signature of Raman-induced mixing between the dressed propagation channels.
Depending on its phase at the exit face, the beating can enhance, suppress, or reverse the extracted third-order rotation.
The eigenchannel analysis in Sec.~\ref{sec:eigenmode} makes this interpretation explicit.

Figures~\ref{fig4}(c) and \ref{fig4}(d) separate the three ingredients of the mechanism.
The EIT and EIT+WM cases, calculated without vdW interactions, show only weak third-order rotations.
Adding vdW interactions without WM dressing gives the EIT+vdW curve, where a finite but still small rotation accumulates.
Only when WM dressing, the EIT-supported Rydberg pathway, and vdW interactions are present together does the third-order rotation become large.
The amplitude-ratio curves in Fig.~\ref{fig4}(d) show that WM dressing is the source of circular-component imbalance.
The vdW interaction can partially reduce that amplitude imbalance, as seen by comparing EIT with EIT+vdW and EIT+WM with EIT+vdW+WM.
This does not contradict the enhancement of \(\psi_{\rm nonlin}\), because the rotation is governed by the accumulated phase and absorption encoded in the density-matrix source terms, not by the amplitude ratio alone.
Thus Fig.~\ref{fig4} identifies the cooperative origin of the giant third-order rotation, while Fig.~\ref{fig5} below diagnoses how the propagation channels participate in that cooperation.

\subsection{Propagation-channel analysis of the third-order rotation}
\label{sec:eigenmode}
\begin{figure*}[htbp]
\centering
\begin{minipage}{0.48\textwidth}
\centering
\includegraphics[width=\textwidth]{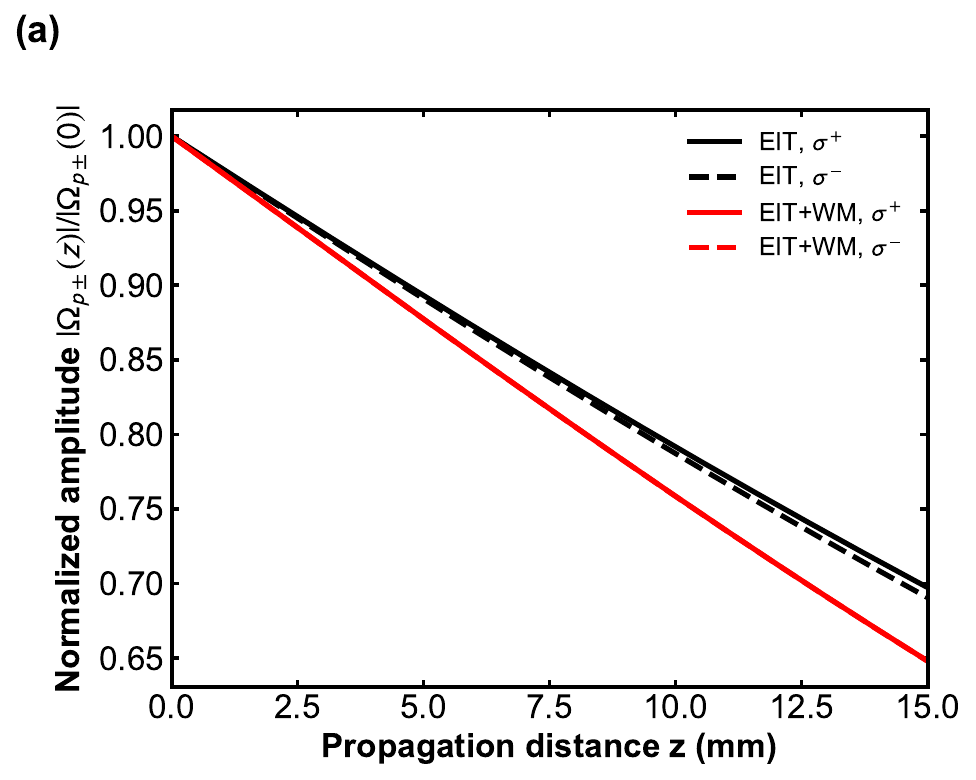}
\end{minipage}
\hfill
\begin{minipage}{0.48\textwidth}
\centering
\includegraphics[width=\textwidth]{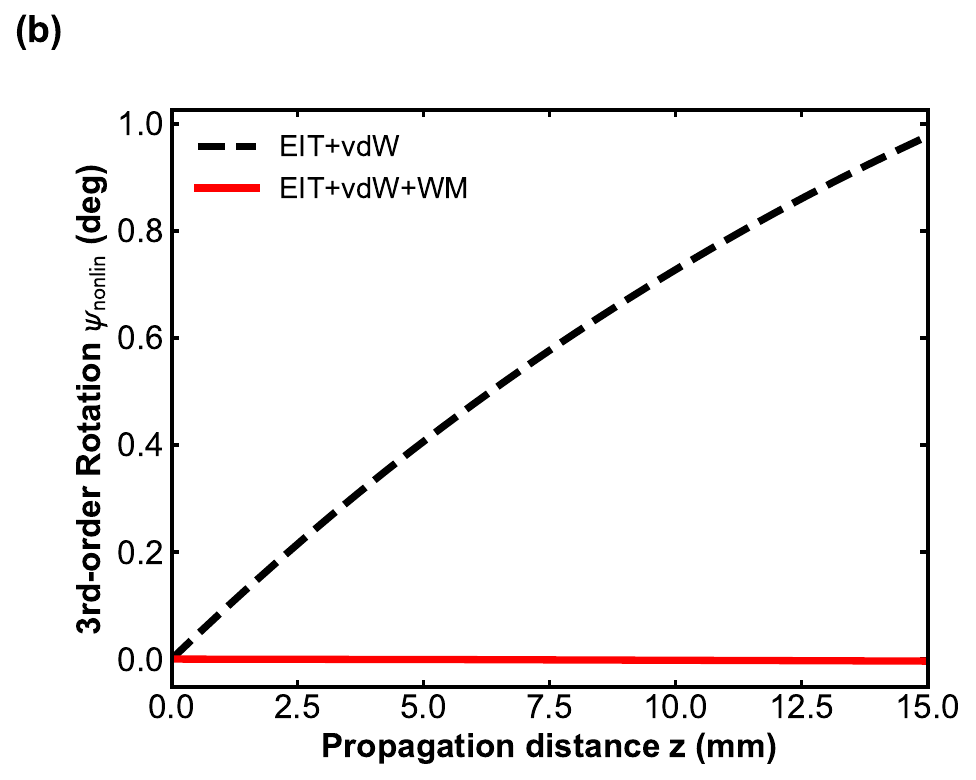}
\end{minipage}
\caption{Propagation-channel diagnostic for the WM-assisted third-order response.
The incident polarization is matched to one WM-dressed linear eigenchannel.
(a) Circular-component amplitudes normalized by their incident magnitudes.
(b) Extracted third-order rotation for the same input.
With WM dressing retained, the selected channel remains nearly isolated and the third-order rotation is strongly suppressed; when the WM field is removed while the same incident polarization is kept, the input is no longer an eigenchannel of the remaining EIT+vdW medium, so differential propagation and the vdW response produce a finite rotation.}
\label{fig5}
\end{figure*}

The WM field mixes the two circular components already in the first-order response.
It is therefore useful to diagonalize the WM-dressed linear propagation problem and ask whether the large third-order rotation is carried by one dressed channel or by the simultaneous propagation of two dressed channels.
This is the purpose of Fig.~\ref{fig5}; it is a diagnostic calculation rather than a new operating configuration.
It rules out the simpler interpretation that the large rotation is carried by one favored dressed channel.

In the weak-probe limit, combining the WM-dressed first-order density-matrix response with Maxwell's equations gives the non-Hermitian linear propagation matrix
\begin{equation}
\partial_z\boldsymbol{\Omega}_p
=
i\mathbf H_{\rm lin}\boldsymbol{\Omega}_p,
\qquad
\boldsymbol{\Omega}_p
=
\begin{pmatrix}
\Omega_{p1}\\
\Omega_{p2}
\end{pmatrix},
\label{eq:linear_propagation_matrix}
\end{equation}
with
\begin{equation}
\mathbf H_{\rm lin}
=
\begin{pmatrix}
H_{11}&H_{12}\\
H_{21}&H_{22}
\end{pmatrix}.
\label{eq:Hlin_components}
\end{equation}
This matrix contains linear dispersion and absorption, the EIT pathway, and the WM-induced Raman cross-coupling. Decay and dephasing make \(\mathbf H_{\rm lin}\) non-Hermitian (see Appendix~\ref{app:propagation_eigenmodes} for the derivation from Appendix~\ref{app:perturbation}).

The complex propagation constants are
\begin{equation}
q_{\pm}
=
\frac{H_{11}+H_{22}}{2}
\pm
\frac{1}{2}
\sqrt{
(H_{11}-H_{22})^2+4H_{12}H_{21}
}.
\label{eq:propagation_eigenvalues}
\end{equation}
Here, \(\operatorname{Re}q_{\pm}\) and \(\operatorname{Im}q_{\pm}\) govern phase accumulation and attenuation, respectively. The splitting \(q_+-q_-\) sets the beating length and differential attenuation between the two dressed channels. The corresponding right eigenvectors are
\begin{equation}
\mathbf v_s
=
\mathcal N_s
\begin{pmatrix}
1\\
r_s
\end{pmatrix},
\qquad
r_s
=
\frac{q_s-H_{11}}{H_{12}},
\qquad s=\pm.
\label{eq:right_eigenvectors}
\end{equation}

Assuming \(\mathbf H_{\rm lin}\) is diagonalizable, a general input can be expanded as
\begin{equation}
\boldsymbol{\Omega}_p(0)
=
c_+\mathbf v_+
+
c_-\mathbf v_-,
\end{equation}
where the coefficients are determined by the biorthogonal left eigenvectors because the matrix is non-Hermitian. The linear field then evolves as
\begin{equation}
\boldsymbol{\Omega}_p^{(1)}(z)
=
c_+e^{iq_+z}\mathbf v_+
+
c_-e^{iq_-z}\mathbf v_-.
\label{eq:linear_two_mode_propagation}
\end{equation}
If both coefficients are nonzero, the output polarization is shaped by relative phase accumulation and differential attenuation of the two channels. By contrast, if the input is matched to one linear eigenchannel,
\begin{equation}
\boldsymbol{\Omega}_p(0)\propto\mathbf v_s
\quad\Longrightarrow\quad
\boldsymbol{\Omega}_p^{(1)}(z)
=
e^{iq_s z}\boldsymbol{\Omega}_p(0),
\qquad
\frac{\Omega_{p2}^{(1)}(z)}{\Omega_{p1}^{(1)}(z)}=r_s.
\label{eq:single_mode_linear_propagation}
\end{equation}
Thus, in the strictly linear problem, an eigenchannel input is a fixed point of the complex amplitude ratio: the whole field only acquires a scalar propagation factor, and the polarization ellipse does not rotate further.
The question tested below is whether the weak third-order response drives the field appreciably away from such a selected channel.

For consistency with the propagation results above, the plotted quantity remains \(\psi_{\rm nonlin}(z)=\psi_{\rm total}(z)-\psi_{\rm lin}(z)\), evaluated from the same Stokes-parameter rotation angle.

In the numerical test, we choose the WM-dressed channel whose input ratio \(r_s\) is closest to the equal-amplitude ratio \(r=1\); the other channel gives the same qualitative conclusion.
Setting \(\Omega_{p2}(0)/\Omega_{p1}(0)=r_s\) selects a single WM-dressed linear propagation channel at the entrance.
Figure~\ref{fig5}(a) shows that, when WM dressing is retained, the two normalized circular-component amplitudes remain almost locked together during full EIT+vdW+WM propagation.
In other words, the third-order terms do not significantly populate the other dressed channel in this weak-probe regime.
Consequently, Fig.~\ref{fig5}(b) shows that the extracted third-order rotation is strongly suppressed.
The comparison without WM is performed without retuning the input polarization: the same complex ratio \(r_s\), selected from the WM-dressed problem, is injected into the EIT+vdW medium.
Once the WM field is removed, this incident state is no longer a fixed point of the remaining linear propagation matrix.
It therefore decomposes into the available propagation channels of the no-WM medium, accumulates differential phase and attenuation, and allows the vdW-induced third-order source to generate a finite rotation.
Thus the reappearing EIT+vdW curve in panel (b) does not mean that WM suppresses the nonlinearity in general; it shows that the suppression occurs only when the input is eigenmatched to a single WM-dressed propagation channel.

The perturbative structure explains why this diagnostic works.
With \(\mathbf x^{(n)}=(\rho_{31}^{(n)},\rho_{32}^{(n)},\rho_{41}^{(n)},\rho_{42}^{(n)})^T\), the first- and third-order four-component single-body coherence subsystems can be written as
\begin{equation}
\mathbf M_{\rm coh}\mathbf x^{(1)}
=
\mathbf S^{(1)},
\qquad
\mathbf M_{\rm coh}\mathbf x^{(3)}
=
\mathbf S^{(3)}.
\label{eq:shared_coherence_matrix}
\end{equation}
The coefficient matrix \(\mathbf M_{\rm coh}\) is the same at both orders, while the source terms are different: \(\mathbf S^{(3)}\) contains lower-order populations, coherences, and vdW-mediated two-body correlations that are absent from \(\mathbf S^{(1)}\). The explicit matrix and source vectors are given in Appendix~\ref{app:propagation_eigenmodes}. Therefore a linear propagation eigenchannel is not a formal nonlinear eigenstate of the full problem. Nevertheless, Fig.~\ref{fig5} shows that, for the weak-probe parameters considered here, an input matched to one WM-dressed linear eigenchannel remains effectively single-channel and acquires almost no additional third-order rotation.

This result gives the intended interpretation of Figs.~\ref{fig4} and \ref{fig5} together.
The large \(\psi_{\rm nonlin}\) in Fig.~\ref{fig4}(c) does not come from an isolated dressed channel.
It comes from coherent simultaneous excitation of both WM-dressed propagation channels by the usual equal-amplitude input.
The two channels then accumulate different complex phases and attenuations along the medium, continuously reshaping the lower-order source terms.
The vdW-mediated third-order response converts this evolving two-channel asymmetry into the giant polarization rotation.

\subsection{Robustness against parameter variations}
\begin{figure*}[htbp]
\centering
\begin{minipage}[t]{0.48\textwidth}
\centering
\makebox[\textwidth][l]{\hspace*{0.035\textwidth}\textbf{(a)}}\\[-0.5ex]
\includegraphics[width=\textwidth]{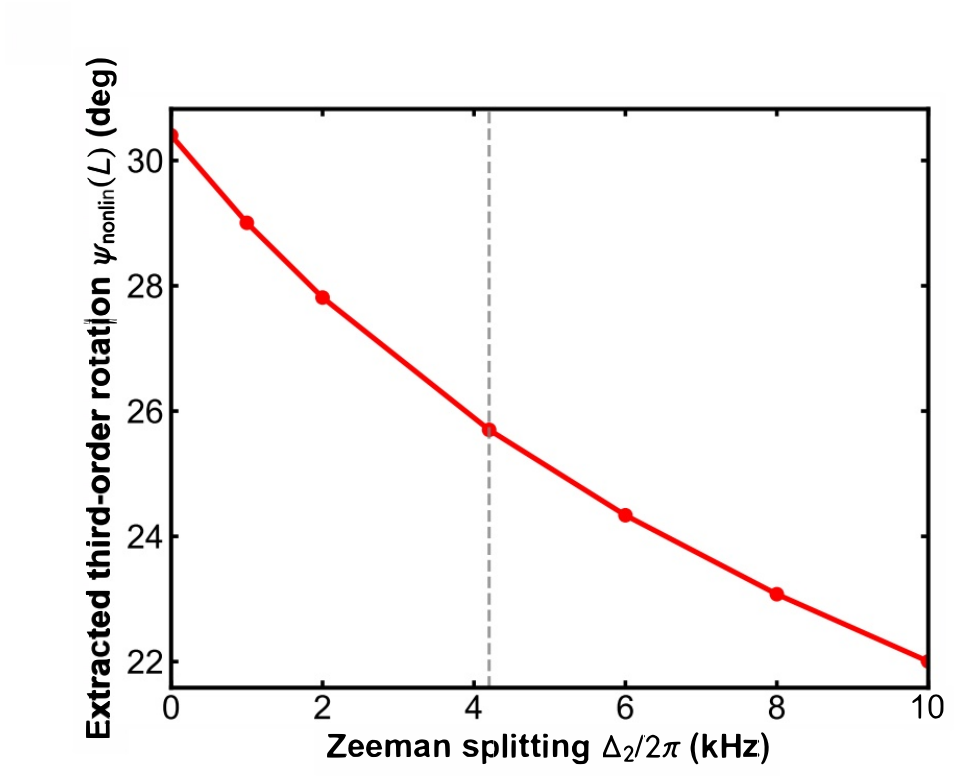}
\end{minipage}
\hfill
\begin{minipage}[t]{0.48\textwidth}
\centering
\makebox[\textwidth][l]{\hspace*{0.035\textwidth}\textbf{(b)}}\\[-0.5ex]
\includegraphics[width=\textwidth]{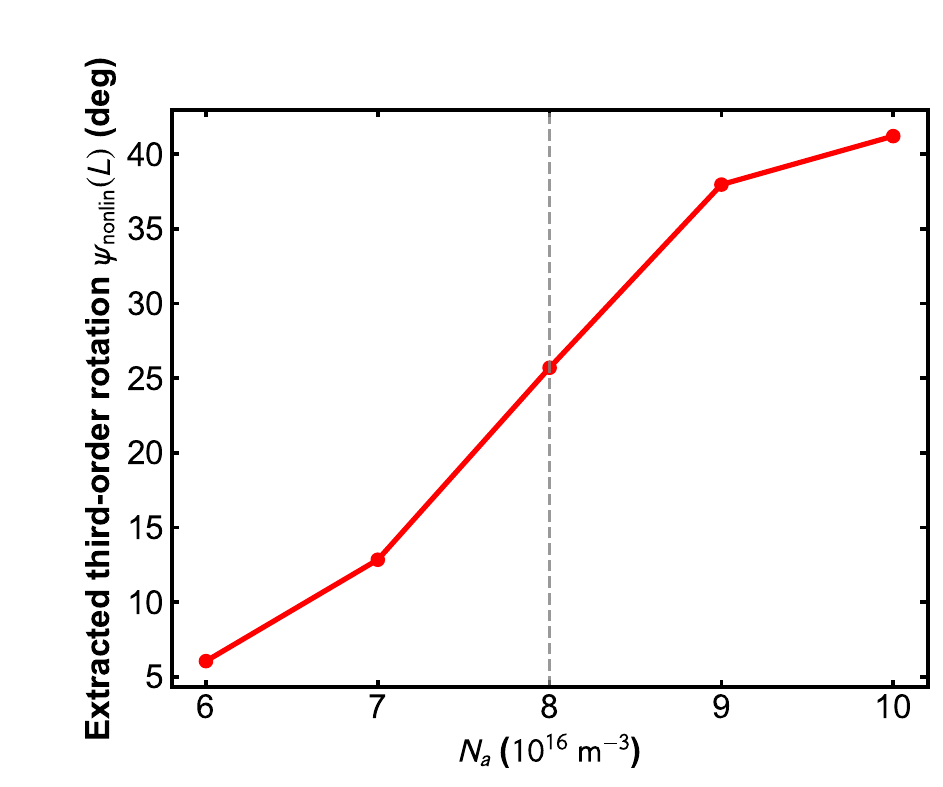}
\end{minipage}
\caption{Robustness of the extracted third-order polarization rotation. (a) Extracted output rotation \(\psi_{\rm nonlin}(L)\) versus the ground-state Zeeman splitting \(\Delta_2/2\pi\), with all other parameters fixed at their nominal values. The dashed marker identifies the nominal operating point \(\Delta_2/2\pi=4.2~\mathrm{kHz}\). The response varies smoothly over the displayed interval and remains finite at zero Zeeman splitting, showing that the nonlinear rotation is not tied to an isolated fine-tuned value of \(\Delta_2\). (b) Extracted output third-order rotation versus atomic density \(\mathcal{N}_a\), with all other parameters fixed at their nominal values. The dotted marker identifies the nominal density \(\mathcal{N}_a=8\times10^{16}\,\mathrm{m}^{-3}\). The response varies continuously over the displayed density interval, showing that the large WM-assisted nonlinear rotation is not tied to an isolated finely tuned density.}
\label{fig6}
\end{figure*}

The ground-state splitting is not a threshold parameter for the nonlinear rotation. Figure~\ref{fig6}(a) shows the extracted output third-order rotation as \(\Delta_2/2\pi\) is varied over \(0,1,2,4.2,6,8,10~\mathrm{kHz}\), with all other parameters fixed at their nominal values. The response changes smoothly from about \(30^{\circ}\) at zero splitting to \(25.70^{\circ}\) at the nominal \(4.2~\mathrm{kHz}\) point and remains sizeable, about \(22^{\circ}\), at \(10~\mathrm{kHz}\). Thus \(4.2~\mathrm{kHz}\) is representative rather than an isolated or finely tuned operating point, and a finite nonlinear rotation persists at zero Zeeman splitting.

The $\Delta_2$ dependence is not a simple additive magnetic-rotation contribution. Changing the splitting modifies the complex detunings, the WM-dressed propagation background, and the lower-order response on which the Rydberg-mediated third-order source evolves. The resulting change in $\psi_{\rm nonlin}$ is therefore a finite-length propagation-accumulation effect; the weak magnetic field is best viewed as a tunable signed propagation bias rather than a prerequisite for the nonlinear rotation.

Figure~\ref{fig6}(b) tests robustness against atomic density. With all other parameters fixed, varying $\mathcal{N}_a$ from $6\times10^{16}$ to $10\times10^{16}\,\mathrm{m}^{-3}$ changes $\psi_{\rm nonlin}(L)$ smoothly from about $6^{\circ}$ to about $41^{\circ}$, with the nominal point lying on the same continuous trend. A separate no-WM calculation confirms that the WM-assisted nonlinear rotation remains substantially larger over the same interval. Because density affects both optical-depth-dependent propagation and the collective Rydberg response, we regard the nominal density as representative and infer no simple scaling law from this limited scan.

We also tested an additional ground-state pure-dephasing rate $\gamma_\phi$, implemented as $D_{21}\rightarrow D_{21}+i\gamma_\phi$ with the population-transfer rates fixed; this standard density-matrix treatment~\cite{ManjappaDelta2014,LiuDephasing2020} leaves the WM-assisted nonlinear rotation at the tens-of-degrees level for $\gamma_\phi/2\pi=0$--$10~\mathrm{kHz}$. The residual-Doppler estimate above gives a conservative crossover scale of several microkelvin for the zero-temperature approximation, rather than a sharp thermal threshold.

\begin{figure*}[t!]
\centering
\begin{minipage}[t]{0.48\textwidth}
\centering
\makebox[\textwidth][l]{\hspace*{0.035\textwidth}\textbf{(a)}}\\[-0.5ex]
\includegraphics[width=\textwidth]{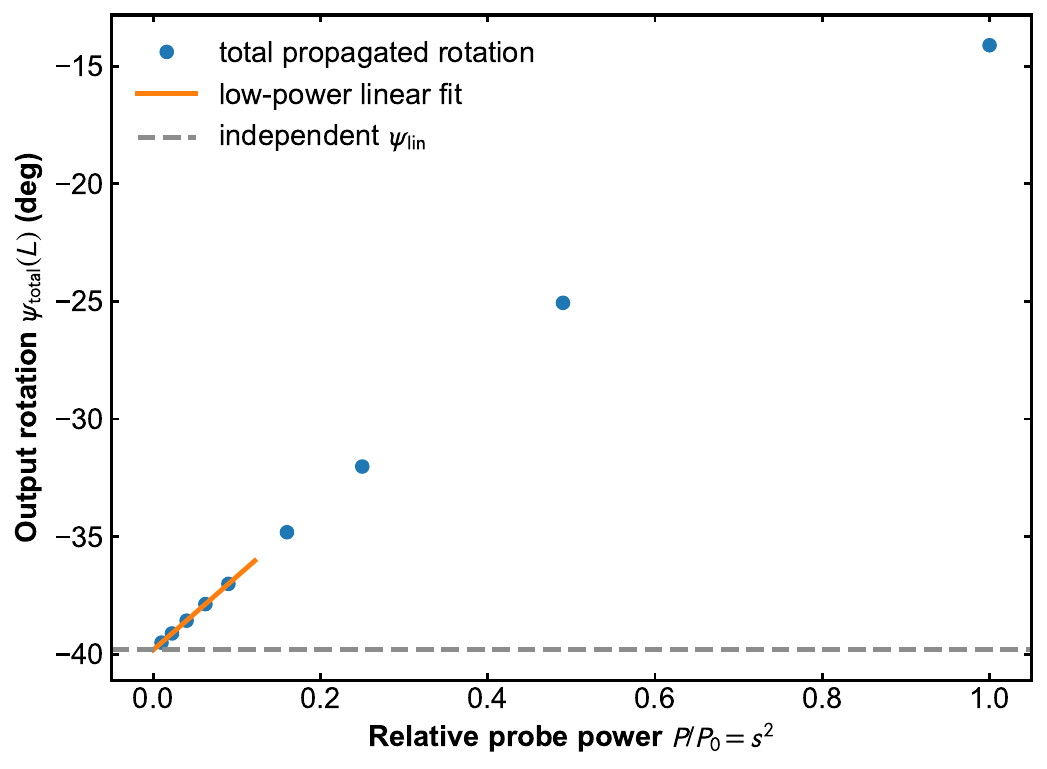}
\end{minipage}
\hfill
\begin{minipage}[t]{0.48\textwidth}
\centering
\makebox[\textwidth][l]{\hspace*{0.035\textwidth}\textbf{(b)}}\\[-0.5ex]
\includegraphics[width=\textwidth]{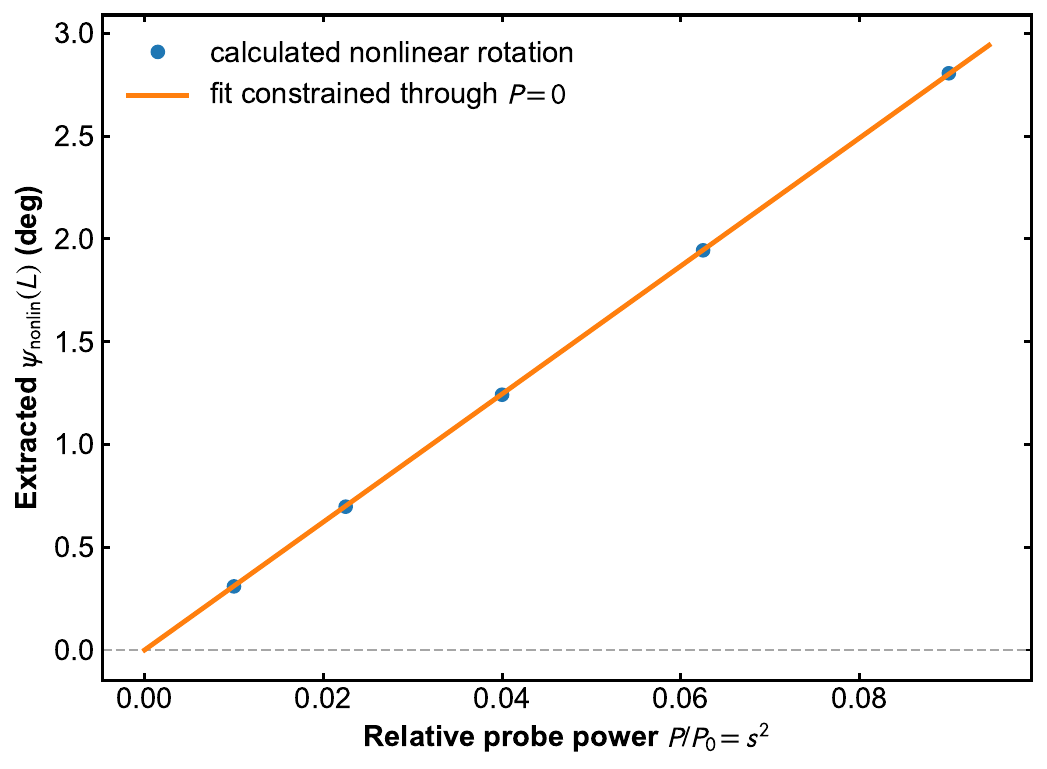}
\end{minipage}
\caption{Probe-power protocol for experimental extraction of the nonlinear rotation. (a) Total propagated output rotation \(\psi_{\rm total}(L)\) versus relative probe power \(P/P_0=s^2\). The solid line is a fit restricted to the low-power points, and the dashed horizontal line is the independently calculated linear background \(\psi_{\rm lin}=-39.81^{\circ}\). The low-power extrapolation approaches the same linear-background value as \(P\rightarrow0\). (b) Extracted nonlinear rotation \(\psi_{\rm nonlin}(L)\) over the low-power interval. The approximately linear dependence on \(P/P_0\), with the fit constrained through the origin, is consistent with the expected cubic weak-probe response.}
\label{fig7}
\end{figure*}

\subsection{Discussion}
Because the vdW-mediated nonlinear response cannot be switched off independently while preserving the same dressed background, we propose a low-power extrapolation. The two incident probe components are scaled together as $\Omega_{pj}(0)\rightarrow s\Omega_{pj}(0)$, so $P/P_0=s^2$, while the medium and all control, WM, magnetic, detuning, and propagation parameters are held fixed.

In the weak-power limit, the first-order polarization is linear in the probe field while the retained third-order term is cubic, so the nonlinear correction to the polarization angle is linear in probe power to leading order:
\[
\psi_{\rm total}(P)=\psi_{\rm lin}+\beta P+O(P^2).
\]
Here $P$ is the physical probe power and the plotted axis uses $P/P_0$. The zero-power intercept of the measured total rotation therefore estimates $\psi_{\rm lin}$ without removing the vdW interaction.

Figure~\ref{fig7}(a) shows that the low-power fit extrapolates to the independently calculated $\psi_{\rm lin}=-39.81^{\circ}$, while Fig.~\ref{fig7}(b) shows an approximately linear low-power dependence of $\psi_{\rm nonlin}$ on $P/P_0$ that vanishes as $P\rightarrow0$. The scaling is consistent with a dominant cubic contribution over the tested weak-power interval. Experimentally, one varies only the incident probe power, obtains the output angle from the measured Stokes parameters, uses the zero-power intercept for $\psi_{\rm lin}$, and subtracts it from the rotation at the chosen operating power.

The results identify a finite-length two-channel mechanism in which WM-induced lower-manifold dressing reshapes the dual-EIT propagation background, while vdW-mediated Rydberg pair correlations supply the microscopic cubic response. Raman-coherence-assisted nonlinear magneto-optical rotation has also been observed in cold rubidium~\cite{ref25}; here the effect is combined with a Rydberg-EIT pathway and an interaction-induced pair response. In the present plane-wave treatment, the spatial pair-correlation contribution is integrated into the local polarization source, so the propagation-level enhancement is governed by the resulting complex two-component response. This viewpoint also suggests that the same propagation principle may extend to other phase-controlled lower-manifold couplings, including Raman or suitably engineered microwave/RF schemes~\cite{WrightRaman2008,LiMicrowaveEIT2009,ManjappaDelta2014}.

The nonmonotonic dependence on $\Omega_{\rm WM}$ and the eigenchannel test provide complementary evidence for this picture: WM strength controls the dressed-channel structure, and the large third-order rotation appears when both dressed channels are coherently excited, whereas eigenmatching to a single channel strongly suppresses the response.

\section{Conclusion}

We have shown that a far-detuned WM field can be used as a Raman dressing background to generate giant third-order polarization rotation in a Rydberg-EIT medium.
After adiabatic elimination, the WM beam is not a probe-gain channel and is not treated as an independent nonlinear source.
Instead, it establishes a Raman-dressed two-channel propagation background whose relative phase and attenuation are tunable by the weak signed Zeeman bias.
The vdW interaction then enters through RDME pair-correlation source terms, and Maxwell-Bloch propagation converts these terms into an accumulated polarimetric signal.

In the benchmark comparison, subtraction of the corresponding linear background gives \(\psi_{\rm nonlin}=+1.06^{\circ}\) for EIT+vdW and \(\psi_{\rm nonlin}=+25.70^{\circ}\) for EIT+vdW+WM.
This is a more than 24-fold enhancement of the extracted third-order rotation.
Mechanism analysis shows that coherent lower-manifold dressing, the EIT-supported Rydberg pathway, and vdW correlations act cooperatively: the dressing, realized here by the far-detuned optical WM field, controls the propagation symmetry, EIT provides the Rydberg coherence pathway, and vdW interactions provide the dominant third-order source.
The residual-Doppler estimate gives a conservative crossover scale of several microkelvin for the zero-temperature approximation, rather than a sharp thermal threshold.

Propagation-channel analysis further shows that the enhanced rotation is a coherent two-channel effect.
Launching a single WM-dressed eigenchannel suppresses the third-order rotation, whereas simultaneous excitation of both dressed channels allows their complex propagation constants to reshape the lower-order source terms and feed the Rydberg-mediated third-order response.
This symmetry-breaking Rydberg nonlinearity provides a route to tunable nonlinear polarization rotation, with applications to weak-light polarimetry and all-optical polarization control.
The low-power extrapolation protocol of Fig.~\ref{fig7} further provides a practical way to separate the nonlinear contribution from the linear dressed background without switching off the vdW interaction.

The enhancement also persists over the parameter variations examined, further indicating that the proposed mechanism is not tied to a single finely tuned operating point.

\begin{acknowledgments}
This work is supported by the Key Special Project of the National Key Research and Development Program of China (Grant Nos. 2025YFF0515201, 2025YFF0515200), the Joint Fund for Quantum Major Research Plan of the National Natural Science Foundation of China (Grant No. U25D8014).
\end{acknowledgments}

\clearpage
\onecolumngrid
\vspace{1em}
\appendix
\section{Derivation of the total Hamiltonian and optical Bloch equations}
\subsection{Single-atom Schr\"{o}dinger and Heisenberg pictures}
Considering the interaction between a single atom and external fields, the single-atom Hamiltonian in the Schr\"{o}dinger picture consists of the unperturbed atomic energy and the electric-dipole interaction:
\begin{align}
\hat{H}_{\text{Single}}^{\text{Sch}} = \sum_{j=1}^{5}E_j|j\rangle\langle j| - \hat{\mathbf{d}} \cdot \mathbf{E}(\mathbf{r},t),
\end{align}
where $E_j = \hbar\omega_j$ is the intrinsic energy of level $|j\rangle$, $\hat{\mathbf{d}}$ is the electric dipole moment operator, and $\mathbf{E}(\mathbf{r},t)$ is the total electric field comprising the probe, control, and wave-mixing fields.
Defining the transition operator $\hat{\sigma}_{\alpha\beta} = |\alpha\rangle\langle\beta|$ and applying a unitary transformation to the Heisenberg picture, the single-atom Hamiltonian is represented as
\begin{align}
\hat{H} &= \sum_{\alpha=1}^{5}E_\alpha\hat{\sigma}_{\alpha\alpha} - \sum_{\alpha\beta}(\mathbf{d}_{\alpha\beta} \cdot \mathbf{E})\hat{\sigma}_{\alpha\beta}(t).
\end{align}
\subsection{Slowly varying operators and phase matching}
To eliminate high-frequency spatial and temporal oscillations of the light field, slowly varying operators $\hat{S}_{\alpha\beta}$ are introduced according to the propagation directions (the wave-mixing and probe fields counterpropagate) and specific coupling relations:
\begin{align}
&\text{Probe fields: } \hat{S}_{31} = |1\rangle\langle 3|e^{-i(k_p z - \omega_p t)}, \quad \hat{S}_{32} = |2\rangle\langle 3|e^{-i(k_p z - \omega_p t)}, \nonumber \\
&\text{Control field: } \hat{S}_{43} = |3\rangle\langle 4|e^{i(k_c z + \omega_c t)}, \nonumber \\
&\text{Wave-mixing fields: } \hat{S}_{51} = |1\rangle\langle 5|e^{i(k_{\rm WM} z + \omega_{\rm WM} t)}, \quad \hat{S}_{52} = |2\rangle\langle 5|e^{i(k_{\rm WM} z + \omega_{\rm WM} t)}, \nonumber \\
&\text{Populations: } \hat{S}_{\alpha\alpha} = |\alpha\rangle\langle\alpha|. \nonumber
\end{align}
\subsection{Rotating-wave approximation and detuning definitions}
Applying the rotating-wave approximation (RWA) in the rotating frame, rapidly oscillating high-frequency terms are discarded to extract the time-independent effective driving components.
Setting the ground state $|1\rangle$ as the energy zero point, the relative detunings for each level are defined as
\begin{align}
&\text{Ground-state Zeeman splitting: } \Delta_2 = -(E_2 - E_1)/\hbar, \nonumber \\
&\text{Probe single-photon detuning: } \Delta_3 = \omega_p - (E_3 - E_1)/\hbar, \nonumber \\
&\text{Rydberg two-photon detuning: } \Delta_4 = (\omega_p + \omega_c) - (E_4 - E_1)/\hbar, \nonumber \\
&\text{WM far-detuning: } \Delta_5 = \omega_{\rm WM} - (E_5 - E_1)/\hbar. \nonumber
\end{align}
\subsection{Total Hamiltonian}
Extending the local light-atom interactions to a macroscopic cold atomic ensemble and incorporating the long-range vdW interaction potential $V(\mathbf{r}'-\mathbf{r}) = -C_6/|\mathbf{r}'-\mathbf{r}|^6$ between Rydberg atoms, the total Hamiltonian $\hat{H}_{\text{total}}$ governing the dynamics of the five-level system is
\begin{equation}
\begin{aligned}[b]
\hat{H}_{\text{total}} &= \mathcal{N}_a\int d^3r \bigg\{-\hbar\sum_{j=2}^{5}\Delta_j\hat{S}_{jj}(\mathbf{r},t) \\*
&\quad -\hbar\big[\Omega_{p1}^*\hat{S}_{31}(\mathbf{r},t) + \Omega_{p2}^*\hat{S}_{32}(\mathbf{r},t) + \Omega_c^*\hat{S}_{43}(\mathbf{r},t) \\*
&\quad + \Omega_{{\rm WM}1}^*\hat{S}_{51}(\mathbf{r},t) + \Omega_{{\rm WM}2}^*\hat{S}_{52}(\mathbf{r},t) + \text{H.c.}\big] \\*
&\quad + \mathcal{N}_a\int d^3r' \hat{S}_{44}(\mathbf{r}',t)\hbar V(\mathbf{r}'-\mathbf{r})\hat{S}_{44}(\mathbf{r},t)\bigg\}.
\end{aligned}
\end{equation}
\subsection{Single-body optical Bloch equations}
Based on the definition $\rho_{\alpha\beta} = \langle\hat{S}_{\alpha\beta}\rangle$, the single-body optical Bloch equations can be logically grouped.
The equations for the diagonal elements (populations) are
\begin{subequations}
\label{eq:OBE_pop}
\begin{align}
0 &= (i\partial_t + i\Gamma_{21})\rho_{11} - i\Gamma_{12}\rho_{22} - i\Gamma_{13}\rho_{33} - i\Gamma_{15}\rho_{55} + \Omega_{p1}^*\rho_{31} - \Omega_{p1}\rho_{13} + \Omega_{{\rm WM}1}^*\rho_{51} - \Omega_{{\rm WM}1}\rho_{15}, \\
0 &= (i\partial_t + i\Gamma_{12})\rho_{22} - i\Gamma_{21}\rho_{11} - i\Gamma_{23}\rho_{33} - i\Gamma_{25}\rho_{55} + \Omega_{p2}^*\rho_{32} - \Omega_{p2}\rho_{23} + \Omega_{{\rm WM}2}^*\rho_{52} - \Omega_{{\rm WM}2}\rho_{25}, \\
0 &= (i\partial_t + i\Gamma_3)\rho_{33} - i\Gamma_{34}\rho_{44} - \Omega_{p1}^*\rho_{31} + \Omega_{p1}\rho_{13} - \Omega_{p2}^*\rho_{32} + \Omega_{p2}\rho_{23} + \Omega_c^*\rho_{43} - \Omega_c\rho_{34}, \\
0 &= (i\partial_t + i\Gamma_4)\rho_{44} - \Omega_c^*\rho_{43} + \Omega_c\rho_{34}, \\
0 &= (i\partial_t + i\Gamma_5)\rho_{55} - \Omega_{{\rm WM}1}^*\rho_{51} + \Omega_{{\rm WM}1}\rho_{15} - \Omega_{{\rm WM}2}^*\rho_{52} + \Omega_{{\rm WM}2}\rho_{25},
\end{align}
\end{subequations}
and for the off-diagonal elements (coherences):
\begin{subequations}
\label{eq:OBE_coh}
\begin{align}
0 &= (i\partial_t + d_{21})\rho_{21} + \Omega_{p2}^*\rho_{31} + \Omega_{{\rm WM}2}^*\rho_{51} - \Omega_{p1}\rho_{23} - \Omega_{{\rm WM}1}\rho_{25}, \\
0 &= (i\partial_t + d_{31})\rho_{31} + \Omega_c^*\rho_{41} + \Omega_{p1}(\rho_{11} - \rho_{33}) + \Omega_{p2}\rho_{21} - \Omega_{{\rm WM}1}\rho_{35}, \\
0 &= (i\partial_t + d_{32})\rho_{32} + \Omega_c^*\rho_{42} + \Omega_{p2}(\rho_{22} - \rho_{33}) + \Omega_{p1}\rho_{12} - \Omega_{{\rm WM}2}\rho_{35}, \\
0 &= (i\partial_t + d_{41})\rho_{41} + \Omega_c\rho_{31} - \Omega_{p1}\rho_{43} - \Omega_{{\rm WM}1}\rho_{45} - \mathcal{N}_a\int d^3r' V(\mathbf{r}'-\mathbf{r})\rho_{44,41}(\mathbf{r}',\mathbf{r},t), \\
0 &= (i\partial_t + d_{42})\rho_{42} + \Omega_c\rho_{32} - \Omega_{p2}\rho_{43} - \Omega_{{\rm WM}2}\rho_{45} - \mathcal{N}_a\int d^3r' V(\mathbf{r}'-\mathbf{r})\rho_{44,42}(\mathbf{r}',\mathbf{r},t), \\
0 &= (i\partial_t + d_{43})\rho_{43} + \Omega_c(\rho_{33} - \rho_{44}) - \Omega_{p1}^*\rho_{41} - \Omega_{p2}^*\rho_{42} - \mathcal{N}_a\int d^3r' V(\mathbf{r}'-\mathbf{r})\rho_{44,43}(\mathbf{r}',\mathbf{r},t), \\
0 &= (i\partial_t + d_{51})\rho_{51} + \Omega_{{\rm WM}1}(\rho_{11} - \rho_{55}) + \Omega_{{\rm WM}2}\rho_{21} - \Omega_{p1}\rho_{53}, \\
0 &= (i\partial_t + d_{52})\rho_{52} + \Omega_{{\rm WM}2}(\rho_{22} - \rho_{55}) + \Omega_{{\rm WM}1}\rho_{12} - \Omega_{p2}\rho_{53}, \\
0 &= (i\partial_t + d_{53})\rho_{53} + \Omega_{{\rm WM}1}\rho_{13} + \Omega_{{\rm WM}2}\rho_{23} - \Omega_{p1}^*\rho_{51} - \Omega_{p2}^*\rho_{52} - \Omega_c\rho_{54}, \\
0 &= (i\partial_t + d_{54})\rho_{54} + \Omega_{{\rm WM}1}\rho_{14} + \Omega_{{\rm WM}2}\rho_{24} - \Omega_c^*\rho_{53} + \mathcal{N}_a\int d^3r' V(\mathbf{r}'-\mathbf{r})\rho_{44,54}(\mathbf{r}',\mathbf{r},t).
\end{align}
\end{subequations}
\section{Steady-state adiabatic elimination and renormalization}
\subsection{Adiabatic approximation and upper-level coherence solutions}
The WM field operates in the far-detuned regime. Setting the single-photon detuning as $\Delta_{{\rm WM}1} = \Delta_5 - \Delta_1 \approx \Delta_{{\rm WM}2} = \Delta_5 - \Delta_2$ and adopting the simplification $\Delta_{{\rm WM}1} \approx \Delta_{{\rm WM}2} \approx \Delta_{\rm WM} \equiv \Delta_5$, the total linearly polarized field is specified by $\Omega_{\rm WM}/2\pi=10\,\mathrm{MHz}$ and decomposed according to Eq.~\eqref{eq:WM_circular_decomposition}. Hence $|\Omega_{{\rm WM}j}|/2\pi=(10/\sqrt{2})\,\mathrm{MHz}$ and the adiabatic parameter is $\epsilon_W=|\Omega_{{\rm WM}j}|/|\Delta_{\rm WM}|=(10/\sqrt{2})/2000\simeq3.54\times10^{-3}\ll1$.
The weak-probe perturbation parameter is simultaneously $\epsilon_p = \Omega_p/\Gamma_3 \approx 0.008 \ll 1$ for $\Omega_p=2\pi\times\SI{0.0477}{MHz}$.
The elimination is an expansion in $\Omega_{\rm WM}/d_5$, whereas the optical response is subsequently ordered in the weak-probe parameter $\epsilon_p$.
The retained coherences $\rho_{5j}$ are the leading eliminated optical coherences of order $\mathcal{O}(\epsilon_W)$.
Mixed probe--WM corrections, such as $\Omega_{p1}\rho_{53} \sim \mathcal{O}(\epsilon_W \epsilon_p^2) \sim 10^{-7}$ (compared to $\sim 10^{-3}$ for the retained terms), are therefore safely omitted in the leading adiabatic solution.
Applying steady-state adiabatic elimination ($\partial_t \rightarrow 0$) to the optical Bloch equations, we extract the first-order quasi-steady-state coherences:
\begin{align}
\rho_{51} &\approx -\frac{\Omega_{{\rm WM}1}}{d_{51}}(\rho_{11} - \rho_{55}) - \frac{\Omega_{{\rm WM}2}}{d_{51}}\rho_{21}, \\
\rho_{52} &\approx -\frac{\Omega_{{\rm WM}2}}{d_{52}}(\rho_{22} - \rho_{55}) - \frac{\Omega_{{\rm WM}1}}{d_{52}}\rho_{12}.
\end{align}
Using the large-detuning approximation $d_{51} \approx d_{52} \equiv d_5 = \Delta_{\rm WM} + i\Gamma_5/2$, together with the algebraic identity $1/d_5^* - 1/d_5 = i\Gamma_5/|d_5|^2$, and neglecting higher-order terms involving $\Omega_{\rm WM}/d_5$ in all denominators, the population of level $|5\rangle$, $\rho_{55}$, is self-consistently evaluated as
\begin{equation}
\rho_{55} = \frac{|\Omega_{{\rm WM}1}|^2\rho_{11} + |\Omega_{{\rm WM}2}|^2\rho_{22} + \Omega_{{\rm WM}1}\Omega_{{\rm WM}2}^*\rho_{12} + \Omega_{{\rm WM}1}^*\Omega_{{\rm WM}2}\rho_{21}}{|d_5|^2}.
\end{equation}
Since the real population is strongly suppressed [$\rho_{55} \sim \mathcal{O}(\epsilon_W^2) \sim 1.25 \times 10^{-5}$], the truncation $(\rho_{jj} - \rho_{55}) \approx \rho_{jj}$ is justified.
The higher-order coherences $\rho_{53}$ and $\rho_{54}$ are derived similarly:
\begin{align}
\rho_{53} &\approx \frac{-d_5(\Omega_{{\rm WM}1}\rho_{13} + \Omega_{{\rm WM}2}\rho_{23}) - \Omega_c(\Omega_{{\rm WM}1}\rho_{14} + \Omega_{{\rm WM}2}\rho_{24})}{d_5^2}, \\
\rho_{54} &\approx \frac{-d_5(\Omega_{{\rm WM}1}\rho_{14} + \Omega_{{\rm WM}2}\rho_{24}) - \Omega_c^*(\Omega_{{\rm WM}1}\rho_{13} + \Omega_{{\rm WM}2}\rho_{23})}{d_5^2}.
\end{align}
\subsection{Renormalization and truncation of higher-order coherence corrections}
Substituting the wave-mixing polarization back into the evolution equations, we define the renormalization parameters as $\Delta_{AC1} = |\Omega_{{\rm WM}1}|^2/d_5^*$, $\Delta_{AC2} = |\Omega_{{\rm WM}2}|^2/d_5^*$, $\Omega_{C12} = \Omega_{{\rm WM}1}\Omega_{{\rm WM}2}^*/d_5^*$, and $\Omega_{C21} = \Omega_{{\rm WM}2}\Omega_{{\rm WM}1}^*/d_5^*$.
The real parts of $\Delta_{AC1,2}$ are ac Stark light shifts that may be absorbed into the effective detunings or compensated by retuning the two-photon resonance, whereas their small imaginary parts describe off-resonant scattering and remain in the complex detunings.
Modifying the detunings accordingly yields $D_{31} = d_{31} + \Delta_{AC1}$, $D_{41} = d_{41} + \Delta_{AC1}$, $D_{32} = d_{32} + \Delta_{AC2}$, $D_{42} = d_{42} + \Delta_{AC2}$, and $D_{21} = d_{21} + \Delta_{AC1} - \Delta_{AC2}^*$.
Substituting the higher-order coherences $\rho_{35}$ and $\rho_{45}$ introduces higher-order dimensionless correction tensors $W_{ij}$ for the control field $\Omega_c$:
\begin{equation}
W_{ij} = \frac{\Omega_{{\rm WM}i}\Omega_{{\rm WM}j}^*}{(d_5)^2} \quad (i,j \in \{1,2\}).
\end{equation}
The tensors $W_{ij}$ are dimensionless corrections to the optical coherence block. Because their absolute magnitude is heavily suppressed ($|W_{ij}| \sim \mathcal{O}(\epsilon_W^2) \sim 1.25 \times 10^{-5} \ll 1$), the diagonal correction $(1 + W_{11}) \approx 1$ holds securely, and off-diagonal cross-couplings (such as $\Omega_c^* W_{12}\rho_{42}$) constitute secondary contributions.
They are therefore discarded as relative $\mathcal{O}(\epsilon_W^2)$ renormalizations of the control-coupled coherence block.
The Raman couplings $\Omega_{C12}$ and $\Omega_{C21}$, by contrast, are the leading effective frequency-scale couplings generated by the eliminated far-detuned field, with magnitude $\mathcal{O}(|\Omega_{\rm WM}|^2/\Delta_{\rm WM})$; they define the dressing Hamiltonian and are retained.
The reduced four-level optical Bloch equations become
\begin{align}
0 &= (i\partial_t + i\Gamma_{21})\rho_{11} - i\Gamma_{12}\rho_{22} - i\Gamma_{13}\rho_{33} - i\Gamma_{15}\rho_{55} \nonumber \\
&\quad + \Omega_{p1}^*\rho_{31} - \Omega_{p1}\rho_{13} + (\Delta_{AC1} - \Delta_{AC1}^*)\rho_{11} + \Omega_{C12}\rho_{12} - \Omega_{C12}^*\rho_{21}, \\
0 &= (i\partial_t + i\Gamma_{12})\rho_{22} - i\Gamma_{21}\rho_{11} - i\Gamma_{23}\rho_{33} - i\Gamma_{25}\rho_{55} \nonumber \\
&\quad + \Omega_{p2}^*\rho_{32} - \Omega_{p2}\rho_{23} + (\Delta_{AC2} - \Delta_{AC2}^*)\rho_{22} + \Omega_{C21}\rho_{21} - \Omega_{C21}^*\rho_{12}, \\
0 &= (i\partial_t + i\Gamma_3)\rho_{33} - i\Gamma_{34}\rho_{44} - \Omega_{p1}^*\rho_{31} + \Omega_{p1}\rho_{13} - \Omega_{p2}^*\rho_{32} + \Omega_{p2}\rho_{23} + \Omega_c^*\rho_{43} - \Omega_c\rho_{34}, \\
0 &= (i\partial_t + i\Gamma_4)\rho_{44} - \Omega_c^*\rho_{43} + \Omega_c\rho_{34}, \\
0 &= (i\partial_t + D_{21})\rho_{21} + \Omega_{p2}^*\rho_{31} - \Omega_{p1}\rho_{23} + \Omega_{C12}\rho_{22} - \Omega_{C21}^*\rho_{11}, \\
0 &= (i\partial_t + D_{31})\rho_{31} + \Omega_c^*\rho_{41} + \Omega_{p1}(\rho_{11} - \rho_{33}) + \Omega_{p2}\rho_{21} + \Omega_{C12}\rho_{32}, \\
0 &= (i\partial_t + D_{32})\rho_{32} + \Omega_c^*\rho_{42} + \Omega_{p2}(\rho_{22} - \rho_{33}) + \Omega_{p1}\rho_{12} + \Omega_{C21}\rho_{31}, \\
0 &= (i\partial_t + D_{41})\rho_{41} + \Omega_c\rho_{31} - \Omega_{p1}\rho_{43} + \Omega_{C12}\rho_{42} - \mathcal{N}_a\int V(\mathbf{r}'-\mathbf{r})\rho_{44,41} d^3r', \\
0 &= (i\partial_t + D_{42})\rho_{42} + \Omega_c\rho_{32} - \Omega_{p2}\rho_{43} + \Omega_{C21}\rho_{41} - \mathcal{N}_a\int V(\mathbf{r}'-\mathbf{r})\rho_{44,42} d^3r', \\
0 &= (i\partial_t + d_{43})\rho_{43} + \Omega_c(\rho_{33} - \rho_{44}) - \Omega_{p1}^*\rho_{41} - \Omega_{p2}^*\rho_{42} - \mathcal{N}_a\int V(\mathbf{r}'-\mathbf{r})\rho_{44,43} d^3r'.
\end{align}
\section{Order-by-order perturbation expansion}
\label{app:perturbation}
Expanding in the dimensionless perturbation parameter $\epsilon_p$ under the steady-state condition ($\partial_t \rightarrow 0$), the equations decompose systematically into distinct perturbation orders.
\subsection{Zeroth-order equations}
\begin{align}
0 &= i\Gamma_{21}\rho_{11}^{(0)} - i\Gamma_{12}\rho_{22}^{(0)} + (\Delta_{AC1} - \Delta_{AC1}^*)\rho_{11}^{(0)} + \Omega_{C12}\rho_{12}^{(0)} - \Omega_{C12}^*\rho_{21}^{(0)} - i\Gamma_{15}\rho_{55}^{(0)}, \\
0 &= i\Gamma_{12}\rho_{22}^{(0)} - i\Gamma_{21}\rho_{11}^{(0)} + (\Delta_{AC2} - \Delta_{AC2}^*)\rho_{22}^{(0)} + \Omega_{C21}\rho_{21}^{(0)} - \Omega_{C21}^*\rho_{12}^{(0)} - i\Gamma_{25}\rho_{55}^{(0)}, \\
0 &= D_{21}\rho_{21}^{(0)} + \Omega_{C12}\rho_{22}^{(0)} - \Omega_{C21}^*\rho_{11}^{(0)}, \\
\rho_{55}^{(0)} &= \frac{|\Omega_{{\rm WM}1}|^2\rho_{11}^{(0)} + |\Omega_{{\rm WM}2}|^2\rho_{22}^{(0)} + \Omega_{{\rm WM}1}\Omega_{{\rm WM}2}^*\rho_{12}^{(0)} + \Omega_{{\rm WM}1}^*\Omega_{{\rm WM}2}\rho_{21}^{(0)}}{|d_5|^2 }.
\end{align}
\subsection{First-order equations}
\begin{align}
0 &= D_{31}\rho_{31}^{(1)} + \Omega_c^*\rho_{41}^{(1)} + \Omega_{p1}\rho_{11}^{(0)} + \Omega_{p2}\rho_{21}^{(0)} + \Omega_{C12}\rho_{32}^{(1)}, \\
0 &= D_{32}\rho_{32}^{(1)} + \Omega_c^*\rho_{42}^{(1)} + \Omega_{p2}\rho_{22}^{(0)} + \Omega_{p1}\rho_{12}^{(0)} + \Omega_{C21}\rho_{31}^{(1)}, \\
0 &= D_{41}\rho_{41}^{(1)} + \Omega_c\rho_{31}^{(1)} + \Omega_{C12}\rho_{42}^{(1)}, \\
0 &= D_{42}\rho_{42}^{(1)} + \Omega_c\rho_{32}^{(1)} + \Omega_{C21}\rho_{41}^{(1)}.
\end{align}
\subsection{Second-order equations}
\begin{align}
0 &= i\Gamma_{21}\rho_{11}^{(2)} - i\Gamma_{12}\rho_{22}^{(2)} - i\Gamma_{13}\rho_{33}^{(2)} + \Omega_{p1}^*\rho_{31}^{(1)} - \Omega_{p1}\rho_{13}^{(1)} + (\Delta_{AC1} - \Delta_{AC1}^*)\rho_{11}^{(2)} \nonumber \\
&\quad + \Omega_{C12}\rho_{12}^{(2)} - \Omega_{C12}^*\rho_{21}^{(2)} - i\Gamma_{15}\rho_{55}^{(2)}, \\
0 &= i\Gamma_{12}\rho_{22}^{(2)} - i\Gamma_{21}\rho_{11}^{(2)} - i\Gamma_{23}\rho_{33}^{(2)} + \Omega_{p2}^*\rho_{32}^{(1)} - \Omega_{p2}\rho_{23}^{(1)} + (\Delta_{AC2} - \Delta_{AC2}^*)\rho_{22}^{(2)} \nonumber \\
&\quad + \Omega_{C21}\rho_{21}^{(2)} - \Omega_{C21}^*\rho_{12}^{(2)} - i\Gamma_{25}\rho_{55}^{(2)}, \\
0 &= i\Gamma_3\rho_{33}^{(2)} - i\Gamma_{34}\rho_{44}^{(2)} - \Omega_{p1}^*\rho_{31}^{(1)} + \Omega_{p1}\rho_{13}^{(1)} - \Omega_{p2}^*\rho_{32}^{(1)} + \Omega_{p2}\rho_{23}^{(1)} + \Omega_c^*\rho_{43}^{(2)} - \Omega_c\rho_{34}^{(2)}, \\
0 &= i\Gamma_4\rho_{44}^{(2)} - \Omega_c^*\rho_{43}^{(2)} + \Omega_c\rho_{34}^{(2)}, \\
0 &= D_{21}\rho_{21}^{(2)} + \Omega_{p2}^*\rho_{31}^{(1)} - \Omega_{p1}\rho_{23}^{(1)} + \Omega_{C12}\rho_{22}^{(2)} - \Omega_{C21}^*\rho_{11}^{(2)}, \\
0 &= d_{43}\rho_{43}^{(2)} + \Omega_c(\rho_{33}^{(2)} - \rho_{44}^{(2)}) - \Omega_{p1}^*\rho_{41}^{(1)} - \Omega_{p2}^*\rho_{42}^{(1)}, \\
\rho_{55}^{(2)} &= \frac{|\Omega_{{\rm WM}1}|^2\rho_{11}^{(2)} + |\Omega_{{\rm WM}2}|^2\rho_{22}^{(2)} + \Omega_{{\rm WM}1}\Omega_{{\rm WM}2}^*\rho_{12}^{(2)} + \Omega_{{\rm WM}1}^*\Omega_{{\rm WM}2}\rho_{21}^{(2)}}{|d_5|^2}.
\end{align}
\subsection{Third-order equations}
\begin{align}
0 &= D_{31}\rho_{31}^{(3)} + \Omega_c^*\rho_{41}^{(3)} + \Omega_{p1}(\rho_{11}^{(2)} - \rho_{33}^{(2)}) + \Omega_{p2}\rho_{21}^{(2)} + \Omega_{C12}\rho_{32}^{(3)}, \\
0 &= D_{32}\rho_{32}^{(3)} + \Omega_c^*\rho_{42}^{(3)} + \Omega_{p2}(\rho_{22}^{(2)} - \rho_{33}^{(2)}) + \Omega_{p1}\rho_{12}^{(2)} + \Omega_{C21}\rho_{31}^{(3)}, \\
0 &= D_{41}\rho_{41}^{(3)} + \Omega_c\rho_{31}^{(3)} - \Omega_{p1}\rho_{43}^{(2)} + \Omega_{C12}\rho_{42}^{(3)} - \mathcal{N}_a\int V(\mathbf{r}'-\mathbf{r})\rho_{44,41}^{(3)} d^3r', \\
0 &= D_{42}\rho_{42}^{(3)} + \Omega_c\rho_{32}^{(3)} - \Omega_{p2}\rho_{43}^{(2)} + \Omega_{C21}\rho_{41}^{(3)} - \mathcal{N}_a\int V(\mathbf{r}'-\mathbf{r})\rho_{44,42}^{(3)} d^3r'.
\end{align}
\section{RDME-based multidimensional closed cascade equations}
\label{app:rdme}
Using the many-body Heisenberg operator derivations and the RDME truncation rule, the localized evolution matrix for the two-body operator $\rho_{\alpha\beta,\mu\nu}^{(n)} = \langle\hat{S}_{\beta\alpha}\hat{S}_{\nu\mu}\rangle^{(n)}$ decouples cleanly into compact matrix forms (with $V \equiv V(\mathbf{r}'-\mathbf{r})$).
To clearly distinguish the intrinsic system evolution from the low-order source drives, we express the equations in the canonical form $M\mathbf{x} = \mathbf{S}$.
\subsection{Second-order two-body equations}
For a specific $\alpha \in \{1,2\}$, the intra-branch variables form a $3 \times 3$ matrix equation:
\begin{equation}
\begin{pmatrix}
2D_{4\alpha}-V & 2\Omega_c & 0 \\
\Omega_c^* & D_{4\alpha,3\alpha} & \Omega_c \\
0 & 2\Omega_c^* & 2D_{3\alpha}
\end{pmatrix}
\begin{pmatrix}
\rho_{4\alpha,4\alpha}^{(2)} \\
\rho_{4\alpha,3\alpha}^{(2)} \\
\rho_{3\alpha,3\alpha}^{(2)}
\end{pmatrix}
=
\begin{pmatrix}
S_{1}^{(\alpha)} \\
S_{2}^{(\alpha)} \\
S_{3}^{(\alpha)}
\end{pmatrix},
\end{equation}
where $D_{\mu,\nu} \equiv D_\mu + D_\nu$ is introduced for brevity, and the source terms, which embed the Raman dressing responsible for circular-component symmetry breaking $\Omega_C^{(\alpha)} \equiv \Omega_{C12}\delta_{1\alpha} + \Omega_{C21}\delta_{2\alpha}$, are given by
\begin{align}
S_{1}^{(\alpha)} &= - 2\Omega_C^{(\alpha)}\rho_{41,42}^{(2)}, \\
S_{2}^{(\alpha)} &= - \Omega_{p1}\rho_{4\alpha}^{(1)}\rho_{1\alpha}^{(0)} - \Omega_{p2}\rho_{4\alpha}^{(1)}\rho_{2\alpha}^{(0)} - \Omega_C^{(\alpha)}(\rho_{42,31}^{(2)} + \rho_{41,32}^{(2)}), \\
S_{3}^{(\alpha)} &= - 2\Omega_{p1}\rho_{3\alpha}^{(1)}\rho_{1\alpha}^{(0)} - 2\Omega_{p2}\rho_{3\alpha}^{(1)}\rho_{2\alpha}^{(0)} - 2\Omega_C^{(\alpha)}\rho_{31,32}^{(2)}.
\end{align}
The cross-branch correlations ($\alpha = 1 \leftrightarrow 2$) independently form a $4 \times 4$ matrix equation:
\begin{equation}
\setlength{\arraycolsep}{3pt}
\begin{pmatrix}
D_{41,42}-V & \Omega_c & \Omega_c & 0 \\
\Omega_c^* & D_{42,31} & 0 & \Omega_c \\
\Omega_c^* & 0 & D_{41,32} & \Omega_c \\
0 & \Omega_c^* & \Omega_c^* & D_{32,31}
\end{pmatrix}
\begin{pmatrix}
\rho_{42,41}^{(2)} \\
\rho_{42,31}^{(2)} \\
\rho_{41,32}^{(2)} \\
\rho_{32,31}^{(2)}
\end{pmatrix}
=
\begin{pmatrix}
S_{c,1} \\
S_{c,2} \\
S_{c,3} \\
S_{c,4}
\end{pmatrix},
\end{equation}
with the corresponding source vector components
\begin{align}
S_{c,1} &= - \Omega_{C12}\rho_{42,42}^{(2)} - \Omega_{C21}\rho_{41,41}^{(2)}, \\
S_{c,2} &= - \Omega_{p1}\rho_{42}^{(1)}\rho_{11}^{(0)} - \Omega_{p2}\rho_{42}^{(1)}\rho_{21}^{(0)} - \Omega_{C12}\rho_{42,32}^{(2)} - \Omega_{C21}\rho_{41,31}^{(2)}, \\
S_{c,3} &= - \Omega_{p1}\rho_{41}^{(1)}\rho_{12}^{(0)} - \Omega_{p2}\rho_{41}^{(1)}\rho_{22}^{(0)} - \Omega_{C12}\rho_{42,32}^{(2)} - \Omega_{C21}\rho_{41,31}^{(2)}, \\
S_{c,4} &= - \Omega_{p1}\rho_{31}^{(1)}\rho_{12}^{(0)} - \Omega_{p2}\rho_{31}^{(1)}\rho_{22}^{(0)} - \Omega_{C12}\rho_{32,32}^{(2)} \nonumber \\
&\quad - \Omega_{p1}\rho_{32}^{(1)}\rho_{11}^{(0)} - \Omega_{p2}\rho_{32}^{(1)}\rho_{21}^{(0)} - \Omega_{C21}\rho_{31,31}^{(2)}.
\end{align}
\subsection{Third-order two-body equations}
To extract the nonlocal integral source $\rho_{44,4j}^{(3)}$ (evaluated for $\alpha \in \{1,2\}$), the eight coupled variables are compactly cast as an $8 \times 8$ matrix equation.
We define the diagonal elements $A_1 \dots A_8$ as follows:
\begin{align}
A_1 &= -(D_{3\alpha}+i\Gamma_3), \quad A_2 = -(D_{3\alpha}+d_{43}), \nonumber \\
A_3 &= -(D_{3\alpha}+i\Gamma_4), \quad A_4 = -(D_{4\alpha}+i\Gamma_3), \nonumber \\
A_5 &= -(D_{4\alpha}+d_{43}-V), \; A_6 = -(D_{4\alpha}+i\Gamma_4-V), \nonumber \\
A_7 &= -(D_{3\alpha}+d_{34}), \quad A_8 = -(D_{4\alpha}+d_{34}). \nonumber
\end{align}
The evolution system $M_3 \vec{\rho}^{(3)} = \vec{S}^{(3)}$ is then
\begin{equation}
\setlength{\arraycolsep}{2.5pt}
\begin{pmatrix}
A_1 & -\Omega_c^* & i\Gamma_{34} & -\Omega_c^* & 0 & 0 & \Omega_c & 0 \\
-\Omega_c & A_2 & \Omega_c & 0 & -\Omega_c^* & 0 & 0 & 0 \\
0 & \Omega_c^* & A_3 & 0 & 0 & -\Omega_c^* & -\Omega_c & 0 \\
-\Omega_c & 0 & 0 & A_4 & -\Omega_c^* & i\Gamma_{34} & 0 & \Omega_c \\
0 & -\Omega_c & 0 & -\Omega_c & A_5 & \Omega_c & 0 & 0 \\
0 & 0 & -\Omega_c & 0 & \Omega_c^* & A_6 & 0 & -\Omega_c \\
\Omega_c^* & 0 & -\Omega_c^* & 0 & 0 & 0 & A_7 & -\Omega_c^* \\
0 & 0 & 0 & \Omega_c^* & 0 & -\Omega_c^* & -\Omega_c & A_8
\end{pmatrix}
\begin{pmatrix}
\rho_{3\alpha,33}^{(3)} \\ \rho_{3\alpha,43}^{(3)} \\ \rho_{3\alpha,44}^{(3)} \\ \rho_{4\alpha,33}^{(3)} \\ \rho_{4\alpha,43}^{(3)} \\ \rho_{4\alpha,44}^{(3)} \\ \rho_{3\alpha,34}^{(3)} \\ \rho_{4\alpha,34}^{(3)}
\end{pmatrix}
=
\begin{pmatrix}
S_1^{(3)} \\ S_2^{(3)} \\ S_3^{(3)} \\ S_4^{(3)} \\ S_5^{(3)} \\ S_6^{(3)} \\ S_7^{(3)} \\ S_8^{(3)}
\end{pmatrix},
\end{equation}
where the excitation source vector components $\vec{S}^{(3)}$ are
\allowdisplaybreaks
\begin{align}
S_1^{(3)} &= \Omega_{C12}\delta_{1\alpha}\rho_{32,33}^{(3)} + \Omega_{C21}\delta_{2\alpha}\rho_{31,33}^{(3)} + \Omega_{p1}\rho_{33}^{(2)}\rho_{1\alpha}^{(0)} + \Omega_{p2}\rho_{33}^{(2)}\rho_{2\alpha}^{(0)} \nonumber \\
&\quad - \Omega_{p1}^*\rho_{3\alpha,31}^{(2)} + \Omega_{p1}\rho_{3\alpha}^{(1)}\rho_{13}^{(1)} - \Omega_{p2}^*\rho_{3\alpha,32}^{(2)} + \Omega_{p2}\rho_{3\alpha}^{(1)}\rho_{23}^{(1)}, \\
S_2^{(3)} &= \Omega_{C12}\delta_{1\alpha}\rho_{32,43}^{(3)} + \Omega_{C21}\delta_{2\alpha}\rho_{31,43}^{(3)} + \Omega_{p1}\rho_{43}^{(2)}\rho_{1\alpha}^{(0)} + \Omega_{p2}\rho_{43}^{(2)}\rho_{2\alpha}^{(0)} \nonumber \\
&\quad - \Omega_{p1}^*\rho_{3\alpha,41}^{(2)} - \Omega_{p2}^*\rho_{3\alpha,42}^{(2)}, \\
S_3^{(3)} &= \Omega_{C12}\delta_{1\alpha}\rho_{32,44}^{(3)} + \Omega_{C21}\delta_{2\alpha}\rho_{31,44}^{(3)} + \Omega_{p1}\rho_{44}^{(2)}\rho_{1\alpha}^{(0)} + \Omega_{p2}\rho_{44}^{(2)}\rho_{2\alpha}^{(0)}, \\
S_4^{(3)} &= \Omega_{C12}\delta_{1\alpha}\rho_{42,33}^{(3)} + \Omega_{C21}\delta_{2\alpha}\rho_{41,33}^{(3)} - \Omega_{p1}^*\rho_{4\alpha,31}^{(2)} + \Omega_{p1}\rho_{4\alpha}^{(1)}\rho_{13}^{(1)} \nonumber \\
&\quad - \Omega_{p2}^*\rho_{4\alpha,32}^{(2)} + \Omega_{p2}\rho_{4\alpha}^{(1)}\rho_{23}^{(1)}, \\
S_5^{(3)} &= \Omega_{C12}\delta_{1\alpha}\rho_{42,43}^{(3)} + \Omega_{C21}\delta_{2\alpha}\rho_{41,43}^{(3)} - \Omega_{p1}^*\rho_{4\alpha,41}^{(2)} - \Omega_{p2}^*\rho_{4\alpha,42}^{(2)}, \\
S_6^{(3)} &= \Omega_{C12}\delta_{1\alpha}\rho_{42,44}^{(3)} + \Omega_{C21}\delta_{2\alpha}\rho_{41,44}^{(3)} - \Omega_{p1}^*\rho_{4\alpha,41}^{(2)} - \Omega_{p2}^*\rho_{4\alpha,42}^{(2)}, \\
S_7^{(3)} &= \Omega_{C12}\delta_{1\alpha}\rho_{32,34}^{(3)} + \Omega_{C21}\delta_{2\alpha}\rho_{31,34}^{(3)} + \Omega_{p1}\rho_{34}^{(2)}\rho_{1\alpha}^{(0)} + \Omega_{p2}\rho_{34}^{(2)}\rho_{2\alpha}^{(0)} \nonumber \\
&\quad + \Omega_{p1}\rho_{14}^{(1)}\rho_{3\alpha}^{(1)} + \Omega_{p2}\rho_{24}^{(1)}\rho_{3\alpha}^{(1)}, \\
S_8^{(3)} &= \Omega_{C12}\delta_{1\alpha}\rho_{42,34}^{(3)} + \Omega_{C21}\delta_{2\alpha}\rho_{41,34}^{(3)} + \Omega_{p1}\rho_{14}^{(1)}\rho_{4\alpha}^{(1)} + \Omega_{p2}\rho_{24}^{(1)}\rho_{4\alpha}^{(1)}.
\end{align}

\section{Derivation of the WM-dressed linear propagation matrix and its eigenmodes}
\label{app:propagation_eigenmodes}

\subsection{Four-dimensional first-order coherence system}
The first-order coherence equations in Appendix~\ref{app:perturbation} involve the four single-body coherences
\begin{equation}
\mathbf x^{(1)}
=
\begin{pmatrix}
\rho_{31}^{(1)}\\
\rho_{32}^{(1)}\\
\rho_{41}^{(1)}\\
\rho_{42}^{(1)}
\end{pmatrix},
\qquad
\boldsymbol{\Omega}_p
=
\begin{pmatrix}
\Omega_{p1}\\
\Omega_{p2}
\end{pmatrix}.
\label{eq:app_first_order_vectors}
\end{equation}
Reading the four equations in the same variable order gives
\begin{equation}
\mathbf M_{\rm coh}\mathbf x^{(1)}
=
-\mathbf B_0\boldsymbol{\Omega}_p,
\label{eq:app_first_order_matrix}
\end{equation}
where
\begin{equation}
\mathbf M_{\rm coh}
=
\begin{pmatrix}
D_{31} & \Omega_{C12} & \Omega_c^* & 0\\
\Omega_{C21} & D_{32} & 0 & \Omega_c^*\\
\Omega_c & 0 & D_{41} & \Omega_{C12}\\
0 & \Omega_c & \Omega_{C21} & D_{42}
\end{pmatrix},
\label{eq:app_coherence_matrix}
\end{equation}
and
\begin{equation}
\mathbf B_0
=
\begin{pmatrix}
\rho_{11}^{(0)} & \rho_{21}^{(0)}\\
\rho_{12}^{(0)} & \rho_{22}^{(0)}\\
0&0\\
0&0
\end{pmatrix}.
\label{eq:app_source_matrix}
\end{equation}
The minus sign in Eq.~\eqref{eq:app_first_order_matrix} follows directly from moving the probe-dependent source terms in the Appendix-C equations to the right-hand side. The effective Raman couplings $\Omega_{C12}$ and $\Omega_{C21}$ enter the left-hand coherence-mixing matrix directly, whereas the WM-generated zeroth-order ground-state coherences $\rho_{12}^{(0)}$ and $\rho_{21}^{(0)}$ enter through the source matrix. Consequently, the cross-polarized linear response is not determined by the Raman matrix elements alone.

\subsection{Two-dimensional first-order atomic response}
Introduce the projection matrix
\begin{equation}
\mathbf P
=
\begin{pmatrix}
1&0&0&0\\
0&1&0&0
\end{pmatrix},
\label{eq:app_projection_matrix}
\end{equation}
which selects the optical coherences that drive the two probe components. Solving Eq.~\eqref{eq:app_first_order_matrix} yields
\begin{equation}
\begin{pmatrix}
\rho_{31}^{(1)}\\
\rho_{32}^{(1)}
\end{pmatrix}
=
-\mathbf P\mathbf M_{\rm coh}^{-1}\mathbf B_0
\boldsymbol{\Omega}_p
\equiv
\mathbf X^{(1)}\boldsymbol{\Omega}_p,
\label{eq:app_linear_response_matrix}
\end{equation}
with
\begin{equation}
\mathbf X^{(1)}
=
-\mathbf P\mathbf M_{\rm coh}^{-1}\mathbf B_0.
\label{eq:app_X1_definition}
\end{equation}
No additional approximation is introduced in Eq.~\eqref{eq:app_X1_definition}; it is the exact matrix solution of the WM-dressed first-order subsystem specified in Appendix~\ref{app:perturbation}.

\subsection{Linear propagation matrix from the Maxwell equations}
Define the field-coupling matrix
\begin{equation}
\mathbf K
=
\begin{pmatrix}
\kappa_1&0\\
0&\kappa_2
\end{pmatrix}.
\label{eq:app_K_matrix}
\end{equation}
In the continuous-wave limit, the two Maxwell equations in the main text are equivalent to
\begin{equation}
\partial_z\boldsymbol{\Omega}_p
=
i\mathbf K
\begin{pmatrix}
\rho_{31}^{(1)}\\
\rho_{32}^{(1)}
\end{pmatrix}
=
i\mathbf K\mathbf X^{(1)}\boldsymbol{\Omega}_p.
\label{eq:app_maxwell_matrix}
\end{equation}
Therefore,
\begin{equation}
{
\mathbf H_{\rm lin}
=
\mathbf K\mathbf X^{(1)}
=
-\mathbf K\mathbf P
\mathbf M_{\rm coh}^{-1}\mathbf B_0
}.
\label{eq:app_Hlin_exact}
\end{equation}
The matrix $\mathbf H_{\rm lin}$ has dimensions of inverse length and combines the atomic first-order response with the field-propagation coupling. The complex detunings $D_{\alpha\beta}$ include ac Stark shifts, decay, and dephasing, so $\mathbf H_{\rm lin}$ is generally non-Hermitian. For a uniform medium with fixed external fields and a fixed WM-dressed weak-probe background, its coefficients are independent of $z$. This fixed linear matrix does not, however, diagonalize the complete third-order propagation problem, whose sources evolve through lower-order populations, coherences, and nonlocal correlations.

\subsection{Characteristic equation and propagation constants}
Writing
\begin{equation}
\det(\mathbf H_{\rm lin}-q\mathbf I)=0
\label{eq:app_characteristic_condition}
\end{equation}
gives
\begin{equation}
\begin{vmatrix}
H_{11}-q&H_{12}\\
H_{21}&H_{22}-q
\end{vmatrix}
=
(H_{11}-q)(H_{22}-q)-H_{12}H_{21}=0.
\label{eq:app_determinant_expanded}
\end{equation}
Hence the characteristic polynomial is
\begin{equation}
q^2-(H_{11}+H_{22})q
+(H_{11}H_{22}-H_{12}H_{21})=0,
\label{eq:app_characteristic_polynomial}
\end{equation}
whose two roots are
\begin{equation}
q_{\pm}
=
\frac{H_{11}+H_{22}}{2}
\pm
\frac{1}{2}
\sqrt{(H_{11}-H_{22})^2+4H_{12}H_{21}},
\label{eq:app_eigenvalues}
\end{equation}
in agreement with Eq.~\eqref{eq:propagation_eigenvalues}.

\subsection{Complex-amplitude ratios of the right eigenvectors}
Let a right eigenvector be written as
\begin{equation}
\mathbf v_s
=
\mathcal N_s
\begin{pmatrix}
1\\
r_s
\end{pmatrix},
\qquad
\mathbf H_{\rm lin}\mathbf v_s=q_s\mathbf v_s.
\label{eq:app_right_eigenvector_ansatz}
\end{equation}
The two rows of the eigenvalue equation are
\begin{align}
H_{11}+H_{12}r_s&=q_s,\label{eq:app_eigenvector_row1}\\
H_{21}+H_{22}r_s&=q_s r_s.\label{eq:app_eigenvector_row2}
\end{align}
Whenever the corresponding denominators are nonzero, these equations give
\begin{equation}
r_s
=
\frac{q_s-H_{11}}{H_{12}}
=
\frac{H_{21}}{q_s-H_{22}}.
\label{eq:app_ratio_forms}
\end{equation}
Eliminating $q_s$ between Eqs.~\eqref{eq:app_eigenvector_row1} and \eqref{eq:app_eigenvector_row2} yields
\begin{equation}
H_{12}r_s^2
+
(H_{11}-H_{22})r_s
-
H_{21}
=
0.
\label{eq:app_ratio_quadratic}
\end{equation}
If $H_{12}=0$, the first expression in Eq.~\eqref{eq:app_ratio_forms} is unavailable, but the second remains valid when $q_s\neq H_{22}$. A pure first circular component corresponds to $r_s=0$ and can occur when $H_{21}=0$ with $q_s=H_{11}$. A pure second circular component is represented by the limiting ratio $r_s\to\infty$, for which one uses the eigenvector $(0,1)^T$; this requires $H_{12}=0$ with $q_s=H_{22}$. These limiting cases are coordinate-basis eigenpolarizations.

\subsection{Fixed points of the complex-amplitude-ratio dynamics}
For a field with $\Omega_{p1}(z)\neq0$, define
\begin{equation}
r(z)
=
\frac{\Omega_{p2}(z)}{\Omega_{p1}(z)}.
\label{eq:app_ratio_definition}
\end{equation}
The two components of Eq.~\eqref{eq:linear_propagation_matrix} are
\begin{align}
\frac{d\Omega_{p1}}{dz}
&=
i\left(H_{11}\Omega_{p1}+H_{12}\Omega_{p2}\right),\label{eq:app_component_prop1}\\
\frac{d\Omega_{p2}}{dz}
&=
i\left(H_{21}\Omega_{p1}+H_{22}\Omega_{p2}\right).\label{eq:app_component_prop2}
\end{align}
Using the quotient rule,
\begin{align}
\frac{dr}{dz}
&=
\frac{1}{\Omega_{p1}}\frac{d\Omega_{p2}}{dz}
-
\frac{\Omega_{p2}}{\Omega_{p1}^2}\frac{d\Omega_{p1}}{dz}\nonumber\\
&=
i\left[H_{21}+(H_{22}-H_{11})r-H_{12}r^2\right].
\label{eq:app_ratio_dynamics_unscaled}
\end{align}
For nonzero $r$, this may be written as
\begin{equation}
\frac{1}{r}\frac{dr}{dz}
=
i\left[
H_{22}-H_{11}
+
\frac{H_{21}}{r}
-
H_{12}r
\right].
\label{eq:app_ratio_dynamics}
\end{equation}
The fixed-point condition $dr/dz=0$ is exactly Eq.~\eqref{eq:app_ratio_quadratic}. Thus, the linear propagation eigenpolarizations are the fixed points of the complex-amplitude-ratio dynamics. Since the polarization-ellipse orientation angle is defined in the main text from the Stokes parameters, a constant complex ratio $r(z)$ means that both the ellipse orientation and ellipticity remain unchanged during linear propagation, even though the overall complex amplitude acquires the factor $e^{iq_s z}$.

\subsection{Biorthogonal expansion for the non-Hermitian propagation matrix}
Assume that $\mathbf H_{\rm lin}$ is diagonalizable. Define right and left eigenvectors by
\begin{equation}
\mathbf H_{\rm lin}\mathbf v_s=q_s\mathbf v_s,
\qquad
\mathbf w_s^\dagger\mathbf H_{\rm lin}
=q_s\mathbf w_s^\dagger,
\qquad s=\pm,
\label{eq:app_left_right_eigenvectors}
\end{equation}
and choose the biorthogonal normalization
\begin{equation}
\mathbf w_s^\dagger\mathbf v_t=\delta_{st}.
\label{eq:app_biorthogonal_normalization}
\end{equation}
The modal amplitudes of an arbitrary input are then
\begin{equation}
c_s
=
\mathbf w_s^\dagger\boldsymbol{\Omega}_p(0),
\label{eq:app_modal_coefficients}
\end{equation}
and the propagated field is
\begin{equation}
\boldsymbol{\Omega}_p(z)
=
\sum_{s=\pm}
c_s e^{iq_s z}\mathbf v_s.
\label{eq:app_biorthogonal_propagation}
\end{equation}
The real part of $q_+-q_-$ produces relative phase beating, whereas its imaginary part produces differential attenuation. A single-eigenmode input sets one of the two coefficients to zero and eliminates the two-mode beating. For all numerical parameter sets considered in this work, the propagation matrix is nondegenerate and its two right eigenvectors form a complete basis.

\subsection{Relation to the third-order coherence equations}
Define the third-order vector in the same variable order,
\begin{equation}
\mathbf x^{(3)}
=
\begin{pmatrix}
\rho_{31}^{(3)}\\
\rho_{32}^{(3)}\\
\rho_{41}^{(3)}\\
\rho_{42}^{(3)}
\end{pmatrix}.
\label{eq:app_third_order_vector}
\end{equation}
The first- and third-order four-component single-body coherence subsystems can then be written with the common coefficient matrix of Eq.~\eqref{eq:app_coherence_matrix}:
\begin{equation}
\mathbf M_{\rm coh}\mathbf x^{(1)}
=
\mathbf S^{(1)},
\qquad
\mathbf M_{\rm coh}\mathbf x^{(3)}
=
\mathbf S^{(3)},
\label{eq:app_shared_coherence_matrix}
\end{equation}
where
\begin{equation}
\mathbf S^{(1)}
=
-\mathbf B_0\boldsymbol{\Omega}_p,
\label{eq:app_first_order_source}
\end{equation}
and, from the third-order equations in Appendix~\ref{app:perturbation},
\begin{equation}
\mathbf S^{(3)}
=
\begin{pmatrix}
-\Omega_{p1}(\rho_{11}^{(2)}-\rho_{33}^{(2)})-\Omega_{p2}\rho_{21}^{(2)}\\
-\Omega_{p2}(\rho_{22}^{(2)}-\rho_{33}^{(2)})-\Omega_{p1}\rho_{12}^{(2)}\\
\Omega_{p1}\rho_{43}^{(2)}+\mathcal N_a\displaystyle\int d^3r'\,V(\mathbf r'-\mathbf r)\rho_{44,41}^{(3)}\\
\Omega_{p2}\rho_{43}^{(2)}+\mathcal N_a\displaystyle\int d^3r'\,V(\mathbf r'-\mathbf r)\rho_{44,42}^{(3)}
\end{pmatrix}.
\label{eq:app_third_order_source}
\end{equation}
The first- and third-order four-component single-body coherence subsystems therefore share the same coefficient matrix. The complete third-order problem is nonetheless not identical to the linear one, because its source vector contains second-order populations, coherences, and vdW-mediated two-body correlations determined by the RDME hierarchy, and those sources vary self-consistently with the propagating fields. It therefore does not follow that a linear propagation eigenmode must be an eigenmode of the full third-order dynamics. The eigenmode-resolved numerical result in Fig.~\ref{fig5} is instead a parameter-dependent observation: for the present model and parameter regime, an isolated linear propagation eigenmode produces only a very small third-order rotation.

\twocolumngrid
\bibliography{references_corrected}

@book{ref1,
  title={{Rydberg} Atoms},
  author={Gallagher, T. F.},
  year={1994},
  publisher={Cambridge University Press},
  address={Cambridge},
  doi={10.1017/CBO9780511524530}
}

@article{ref2,
  title={Quantum information with {Rydberg} atoms},
  author={Saffman, M. and Walker, T. G. and M{\o}lmer, K.},
  journal={Rev. Mod. Phys.},
  volume={82},
  pages={2313},
  year={2010},
  doi={10.1103/RevModPhys.82.2313}
}

@article{ref3,
  title={{Rydberg} atom quantum technologies},
  author={Adams, C. S. and Pritchard, J. D. and Shaffer, J. P.},
  journal={J. Phys. B: At. Mol. Opt. Phys.},
  volume={53},
  pages={012002},
  year={2019},
  doi={10.1088/1361-6455/ab52ef}
}

@article{ref4,
  title={Coherent optical detection of highly excited {Rydberg} states using electromagnetically induced transparency},
  author={Mohapatra, A. K. and Jackson, T. R. and Adams, C. S.},
  journal={Phys. Rev. Lett.},
  volume={98},
  pages={113003},
  year={2007},
  doi={10.1103/PhysRevLett.98.113003}
}

@article{ref5,
  title={Dipole blockade and quantum information processing in mesoscopic atomic ensembles},
  author={Lukin, M. D. and Fleischhauer, M. and C{\^o}t{\'e}, R. and Duan, L. M. and Jaksch, D. and Cirac, J. I. and Zoller, P.},
  journal={Phys. Rev. Lett.},
  volume={87},
  pages={037901},
  year={2001},
  doi={10.1103/PhysRevLett.87.037901}
}

@article{ref6,
  title={Local blockade of {Rydberg} excitation in an ultracold gas},
  author={Tong, D. and Farooqi, S. M. and Stanojevic, J. and Krishnan, S. and Zhang, Y. P. and C{\^o}t{\'e}, R. and Eyler, E. E. and Gould, P. L.},
  journal={Phys. Rev. Lett.},
  volume={93},
  pages={063001},
  year={2004},
  doi={10.1103/PhysRevLett.93.063001}
}

@article{ref7,
  title={Suppression of excitation and spectral broadening induced by interactions in a cold gas of {Rydberg} atoms},
  author={Singer, K. and Reetz-Lamour, M. and Amthor, T. and Marcassa, L. G. and Weidem{\"u}ller, M.},
  journal={Phys. Rev. Lett.},
  volume={93},
  pages={163001},
  year={2004},
  doi={10.1103/PhysRevLett.93.163001}
}

@article{ref8,
  title={Observation of {Rydberg} blockade between two atoms},
  author={Urban, E. and Johnson, T. A. and Henage, T. and Isenhower, L. and Yavuz, D. D. and Walker, T. G. and Saffman, M.},
  journal={Nat. Phys.},
  volume={5},
  pages={110--114},
  year={2009},
  doi={10.1038/nphys1178}
}

@article{ref9,
  title={Attractive photons in a quantum nonlinear medium},
  author={Firstenberg, O. and Peyronel, T. and Liang, Q.-Y. and Gorshkov, A. V. and Lukin, M. D. and Vuleti{\'c}, V.},
  journal={Nature},
  volume={502},
  pages={71},
  year={2013},
  doi={10.1038/nature12512}
}

@article{ref10,
  title={Observation of a large, resonant, cross-{Kerr} nonlinearity in a cold {Rydberg} gas},
  author={Sinclair, J. and Angulo, D. and Lupu-Gladstein, N. and Bonsma-Fisher, K. and Steinberg, A. M.},
  journal={Phys. Rev. Research},
  volume={1},
  pages={033193},
  year={2019},
  doi={10.1103/PhysRevResearch.1.033193}
}

@article{ref11,
  title={Photon-photon interactions via {Rydberg} blockade},
  author={Gorshkov, A. V. and Otterbach, J. and Fleischhauer, M. and Pohl, T. and Lukin, M. D.},
  journal={Phys. Rev. Lett.},
  volume={107},
  pages={133602},
  year={2011},
  doi={10.1103/PhysRevLett.107.133602}
}

@article{ref12,
  title={Single-photon switch based on {Rydberg} blockade},
  author={Baur, S. and Tiarks, D. and Rempe, G. and D\"{u}rr, S.},
  journal={Phys. Rev. Lett.},
  volume={112},
  pages={073901},
  year={2014},
  doi={10.1103/PhysRevLett.112.073901}
}

@article{ref13,
  title={Single-photon transistor using a {F\"{o}rster} resonance},
  author={Tiarks, D. and Baur, S. and Schneider, K. and D{\"u}rr, S. and Rempe, G.},
  journal={Phys. Rev. Lett.},
  volume={113},
  pages={053602},
  year={2014},
  doi={10.1103/PhysRevLett.113.053602}
}

@article{ref14,
  title={Single-photon transistor mediated by interstate {Rydberg} interactions},
  author={Gorniaczyk, H. and Tresp, C. and Schmidt, J. and Fedder, H. and Hofferberth, S.},
  journal={Phys. Rev. Lett.},
  volume={113},
  pages={053601},
  year={2014},
  doi={10.1103/PhysRevLett.113.053601}
}

@article{ref15,
  title={Quantum nonlinear optics with single photons enabled by strongly interacting atoms},
  author={Peyronel, T. and Firstenberg, O. and Liang, Q.-Y. and Hofferberth, S. and Gorshkov, A. V. and Pohl, T. and Lukin, M. D. and Vuleti{\'c}, V.},
  journal={Nature},
  volume={488},
  pages={57},
  year={2012},
  doi={10.1038/nature11361}
}

@article{ref16,
  title={Demonstration of a neutral atom controlled-{NOT} quantum gate},
  author={Isenhower, L. and Urban, E. and Zhang, X. L. and Gill, A. T. and Henage, T. and Johnson, T. A. and Walker, T. G. and Saffman, M.},
  journal={Phys. Rev. Lett.},
  volume={104},
  pages={010503},
  year={2010},
  doi={10.1103/PhysRevLett.104.010503}
}

@article{ref17,
  title={High-fidelity control and entanglement of {Rydberg}-atom qubits},
  author={Levine, H. and Keesling, A. and Omran, A. and Bernien, H. and Schwartz, S. and Zibrov, A. S. and Endres, M. and Greiner, M. and Vuleti{\'c}, V. and Lukin, M. D.},
  journal={Phys. Rev. Lett.},
  volume={121},
  pages={123603},
  year={2018},
  doi={10.1103/PhysRevLett.121.123603}
}

@article{ref18,
  title={Resonant nonlinear magneto-optical effects in atoms},
  author={Budker, D. and Gawlik, W. and Kimball, D. F. and Rochester, S. M. and Yashchuk, V. V. and Weis, A.},
  journal={Rev. Mod. Phys.},
  volume={74},
  pages={1153},
  year={2002},
  doi={10.1103/RevModPhys.74.1153}
}

@article{ref19,
  title={A subfemtotesla multichannel atomic magnetometer},
  author={Kominis, I. K. and Kornack, T. W. and Allred, J. C. and Romalis, M. V.},
  journal={Nature},
  volume={422},
  pages={596},
  year={2003},
  doi={10.1038/nature01484}
}

@article{ref20,
  title={Optical magnetometry},
  author={Budker, D. and Romalis, M.},
  journal={Nat. Phys.},
  volume={3},
  pages={227},
  year={2007},
  doi={10.1038/nphys566}
}

@article{ref21,
  title={Nonlinear magneto-optical rotation with modulated light in tilted magnetic fields},
  author={Pustelny, S. and Rochester, S. M. and Jackson Kimball, D. F. and Yashchuk, V. V. and Budker, D. and Gawlik, W.},
  journal={Phys. Rev. A},
  volume={74},
  pages={063420},
  year={2006},
  doi={10.1103/PhysRevA.74.063420}
}

@article{ref22,
  title={Analysis of atomic magnetometry using metasurface optics for balanced polarimetry},
  author={Yang, Xuting and Benelajla, Meryem and Carpenter, Steven and Choy, Jennifer T.},
  journal={Opt. Express},
  volume={31},
  number={8},
  pages={13436--13446},
  year={2023},
  doi={10.1364/OE.486311}
}

@article{ref23,
  title={Integrated optical rotation detection scheme for chip-scale atomic magnetometer empowered by silicon-rich {SiN$_x$} metalens},
  author={Hu, Jinsheng and Liang, Zihua and Zhou, Peng and Liu, Lu and Hu, Gen and Du, Pengcheng and Ye, Mao},
  journal={Opt. Lett.},
  volume={49},
  number={12},
  pages={3364--3367},
  year={2024},
  doi={10.1364/OL.527932}
}

@article{ref24,
  title={Machine learning assisted vector atomic magnetometry},
  author={Meng, Xin and Zhang, Youwei and Zhang, Xichang and Jin, Shenchao and Wang, Tingran and Jiang, Liang and Xiao, Liantuan and Jia, Suotang and Xiao, Yanhong},
  journal={Nat. Commun.},
  volume={14},
  pages={6105},
  year={2023},
  doi={10.1038/s41467-023-41676-x}
}

@article{ref25,
  title={Comparative study of {Raman}-coherence-assisted nonlinear magneto-optical rotation in {D1} and {D2} transitions using cold rubidium atoms},
  author={Yin, Yiran and Xu, An-Ning and Liu, Bei and Deng, Lu},
  journal={Opt. Lett.},
  volume={51},
  number={6},
  pages={1504--1507},
  year={2026},
  publisher={Optica Publishing Group},
  doi={10.1364/OL.579150}
}

@article{WangEITPR2006,
  title={Controlling the polarization rotation of an optical field via asymmetry in electromagnetically induced transparency},
  author={Wang, Bo and Li, Shujing and Ma, Jie and Wang, Hai and Peng, K. C. and Xiao, Min},
  journal={Phys. Rev. A},
  volume={73},
  pages={051801(R)},
  year={2006},
  doi={10.1103/PhysRevA.73.051801}
}

@article{LiEITPR2006,
  title={Controlled polarization rotation of an optical field in multi-{Zeeman}-sublevel atoms},
  author={Li, Shujing and Wang, Bo and Yang, Xudong and Han, Yanxu and Wang, Hai and Xiao, Min and Peng, K. C.},
  journal={Phys. Rev. A},
  volume={74},
  pages={033821},
  year={2006},
  doi={10.1103/PhysRevA.74.033821}
}

@article{HomboEIPR2012,
  title={Electromagnetically induced polarization rotation in {Na} vapor},
  author={Hombo, Naoya and Taniguchi, Shuya and Sugimura, Shota and Fujita, Kazuki and Mitsunaga, Masaharu},
  journal={J. Opt. Soc. Am. B},
  volume={29},
  number={7},
  pages={1717--1721},
  year={2012},
  doi={10.1364/JOSAB.29.001717}
}

@article{DasPREIT2018,
  title={Polarization rotation with electromagnetically induced transparency in a {V}-type configuration of {Rb} {D1} and {D2} transitions},
  author={Das, Arpita and Das, Bankim Chandra and Bhattacharyya, Dipankar and Chakrabarti, Shrabana and De, Sankar},
  journal={J. Phys. B: At. Mol. Opt. Phys.},
  volume={51},
  pages={175502},
  year={2018},
  doi={10.1088/1361-6455/aad6d5}
}

@article{ChoOIF2005,
  title={Optically induced {Faraday} effect using three-level atoms},
  author={Cho, D. and Choi, Jai Min and Kim, Jang Myun and Park, Q-Han},
  journal={Phys. Rev. A},
  volume={72},
  pages={023821},
  year={2005},
  doi={10.1103/PhysRevA.72.023821}
}

@article{ZhengMFP2024,
  title={Magnetic-free polarization rotation in an atomic vapor cell},
  author={Zheng, Jiang-Bin and Chai, De-Kang and Wang, Zhu-Bo and Chen, Guang-Jie and Hu, Ya-Dong and Chen, Liang and Fan, Hong-Jie and Zhang, Yan-Lei and Dong, Chun-Hua and Zou, Chang-Ling and Guo, Guang-Can and Ye, Ming-Yong and Lin, Gong-Wei and Lin, Xiu-Min},
  journal={Opt. Express},
  volume={32},
  number={1},
  pages={313--324},
  year={2024},
  doi={10.1364/OE.510933}
}

@article{ref26,
  title={Symmetry-breaking inelastic wave-mixing atomic magnetometry},
  author={Zhou, F. and Zhu, C. J. and Hagley, E. W. and Deng, L.},
  journal={Sci. Adv.},
  volume={3},
  pages={e1700422},
  year={2017},
  doi={10.1126/sciadv.1700422}
}

@article{ref27,
  title={Breaking the energy-symmetry-based propagation growth blockade in magneto-optical rotation},
  author={Zhu, C. and Zhou, F. and Zhu, E. Y. and Hagley, E. W. and Deng, L.},
  journal={Phys. Rev. Applied},
  volume={10},
  pages={064013},
  year={2018},
  doi={10.1103/PhysRevApplied.10.064013}
}

@article{ref28,
  title={Magneto-optical rotation: accurate approximated analytical solutions for single-probe atomic magnetometers},
  author={Deng, Lu and Deng, Claire},
  journal={Opt. Express},
  volume={30},
  number={10},
  pages={17392--17399},
  year={2022},
  doi={10.1364/OE.456252}
}

@article{ref29,
  title={Switch and phase shift of photon polarization qubits via double {Rydberg} electromagnetically induced transparency},
  author={Ou, Yao and Huang, Guoxiang},
  journal={Phys. Rev. A},
  volume={109},
  pages={023508},
  year={2024},
  doi={10.1103/PhysRevA.109.023508}
}

@article{ref30,
  title={Nonlocal {Rydberg} enhancement for four-wave-mixing biphoton generation},
  author={Zhao, Hui-Min and Zhang, Xiao-Jun and Artoni, M. and La Rocca, G. C. and Wu, Jin-Hui},
  journal={Phys. Rev. A},
  volume={109},
  pages={043711},
  year={2024},
  doi={10.1103/PhysRevA.109.043711}
}

@article{ref31,
  title={Electromagnetically-induced-transparency spectra of {Rydberg} atoms dressed with dual-tone radio-frequency fields},
  author={Jayaseelan, Maitreyi and Rotunno, Andrew P. and Prajapati, Nikunjkumar and Berweger, Samuel and Artusio-Glimpse, Alexandra B. and Simons, Matthew T. and Holloway, Christopher L.},
  journal={Phys. Rev. A},
  volume={108},
  pages={033712},
  year={2023},
  doi={10.1103/PhysRevA.108.033712}
}

@article{ref32,
  title={Strong microwave-induced cross-{Kerr} effect with {Rydberg} atoms at telecommunication wavelength},
  author={Li, Wenfang and Lam, Mark and Du, Jinjin},
  journal={Phys. Rev. Applied},
  volume={23},
  pages={054011},
  year={2025},
  doi={10.1103/PhysRevApplied.23.054011}
}

@article{ref33,
  title={Giant {Kerr} nonlinearities and magneto-optical rotations in a {Rydberg}-atom gas via double electromagnetically induced transparency},
  author={Mu, Y. and Qin, L. and Shi, Z. and Huang, G.},
  journal={Phys. Rev. A},
  volume={103},
  pages={043709},
  year={2021},
  doi={10.1103/PhysRevA.103.043709}
}

@article{ref34,
  title={Enhanced third-order and fifth-order {Kerr} nonlinearities in a cold atomic system via {Rydberg}-{Rydberg} interaction},
  author={Bai, Z. and Huang, G.},
  journal={Opt. Express},
  volume={24},
  pages={4442},
  year={2016},
  doi={10.1364/OE.24.004442}
}

@article{ref35,
  title={Stable single light bullets and vortices and their active control in cold {Rydberg} gases},
  author={Bai, Zhengyang and Li, Weibin and Huang, Guoxiang},
  journal={Optica},
  volume={6},
  number={3},
  pages={309--317},
  year={2019},
  doi={10.1364/OPTICA.6.000309}
}

@article{ref36,
  title={Coherent population trapping with controlled interparticle interactions},
  author={Schempp, H. and G{\"u}nter, G. and Hofmann, C. S. and Giese, C. and Saliba, S. D. and DePaola, B. D. and Amthor, T. and Weidem{\"u}ller, M. and Sevin\c{c}li, S. and Pohl, T.},
  journal={Phys. Rev. Lett.},
  volume={104},
  pages={173602},
  year={2010},
  doi={10.1103/PhysRevLett.104.173602}
}

@article{ref37,
  title={Quantum interference in interacting three-level {Rydberg} gases: Coherent population trapping and electromagnetically induced transparency},
  author={Sevin\c{c}li, S. and Ates, C. and Pohl, T. and Schempp, H. and Hofmann, C. S. and G{\"u}nter, G. and Amthor, T. and Weidem{\"u}ller, M. and Pritchard, J. D. and Maxwell, D. and Gauguet, A. and Weatherill, K. J. and Jones, M. P. A. and Adams, C. S.},
  journal={J. Phys. B: At. Mol. Opt. Phys.},
  volume={44},
  number={18},
  pages={184018},
  year={2011},
  doi={10.1088/0953-4075/44/18/184018}
}

@article{ref38,
  title={Electromagnetically induced transparency in strongly interacting {Rydberg} gases},
  author={Ates, C. and Sevin\c{c}li, S. and Pohl, T.},
  journal={Phys. Rev. A},
  volume={83},
  pages={041802(R)},
  year={2011},
  doi={10.1103/PhysRevA.83.041802}
}

@article{ref39,
  title={Electromagnetically induced transparency: Optics in coherent media},
  author={Fleischhauer, M. and Imamoglu, A. and Marangos, J. P.},
  journal={Rev. Mod. Phys.},
  volume={77},
  pages={633},
  year={2005},
  doi={10.1103/RevModPhys.77.633}
}

@article{ref40,
  title={Electromagnetically induced transparency and its dispersion properties in a four-level inverted-{Y} atomic system},
  author={Joshi, A. and Xiao, M.},
  journal={Phys. Lett. A},
  volume={317},
  pages={370--377},
  year={2003},
  doi={10.1016/j.physleta.2003.09.010}
}

@article{ref41,
  title={Electromagnetically induced transparency in ladder-type inhomogeneously broadened media: Theory and experiment},
  author={Gea-Banacloche, J. and Li, Y. Q. and Jin, S. Z. and Xiao, M.},
  journal={Phys. Rev. A},
  volume={51},
  pages={576},
  year={1995},
  doi={10.1103/PhysRevA.51.576}
}

@book{ref42,
  title={Nonlinear Optics},
  edition={3rd},
  author={Boyd, R. W.},
  year={2008},
  publisher={Academic},
  address={New York}
}

@article{ref43,
  title={Long-range interactions between alkali {Rydberg} atom pairs correlated to the $ns$--$ns$, $np$--$np$, and $nd$--$nd$ asymptotes},
  author={Singer, K. and Stanojevic, J. and Weidem{\"u}ller, M. and C{\^o}t{\'e}, R.},
  journal={J. Phys. B: At. Mol. Opt. Phys.},
  volume={38},
  pages={S295--S307},
  year={2005},
  doi={10.1088/0953-4075/38/2/021}
}

@article{ref44,
  title={{ARC}: An open-source library for calculating properties of alkali {Rydberg} atoms},
  author={\v{S}ibali\'{c}, N. and Pritchard, J. D. and Adams, C. S. and Weatherill, K. J.},
  journal={Comput. Phys. Commun.},
  volume={220},
  pages={319--331},
  year={2017},
  doi={10.1016/j.cpc.2017.06.015}
}

@article{ref45,
  title={Cooperative Atom-Light Interaction in a Blockaded {Rydberg} Ensemble},
  author={Pritchard, J. D. and Maxwell, D. and Gauguet, A. and Weatherill, K. J. and Jones, M. P. A. and Adams, C. S.},
  journal={Phys. Rev. Lett.},
  volume={105},
  pages={193603},
  year={2010},
  doi={10.1103/PhysRevLett.105.193603}
}

@article{ref46,
  title={Effective operator formalism for open quantum systems},
  author={Reiter, F. and S{\o}rensen, A. S.},
  journal={Phys. Rev. A},
  volume={85},
  pages={032111},
  year={2012},
  doi={10.1103/PhysRevA.85.032111}
}

@article{ref47,
  title={Adiabatic elimination and subspace evolution of open quantum systems},
  author={Finkelstein-Shapiro, D. and Viennot, D. and Saideh, I. and Hansen, T. and Pullerits, T. and Keller, A.},
  journal={Phys. Rev. A},
  volume={101},
  pages={042102},
  year={2020},
  doi={10.1103/PhysRevA.101.042102}
}

@article{ref48,
  title={Stable two-dimensional dark-soliton molecules in {Rydberg} atomic gases},
  author={Yu, Zewei and Hang, Chao},
  journal={Phys. Rev. A},
  volume={113},
  pages={063504},
  year={2026},
  doi={10.1103/z515-d5rt}
}

@article{ref49,
  title={Dispersive optical nonlinearities in a {Rydberg} electromagnetically-induced-transparency medium},
  author={Stanojevic, J. and Parigi, V. and Bimbard, E. and Ourjoumtsev, A. and Grangier, P.},
  journal={Phys. Rev. A},
  volume={88},
  pages={053845},
  year={2013},
  doi={10.1103/PhysRevA.88.053845}
}

@book{ref50,
  title={Principles of Optics},
  edition={7th},
  author={Born, M. and Wolf, E.},
  year={1999},
  publisher={Cambridge University Press},
  address={Cambridge}
}

@book{ref51,
  title={Ellipsometry and Polarized Light},
  author={Azzam, R. M. A. and Bashara, N. M.},
  year={1977},
  publisher={North-Holland},
  address={Amsterdam}
}

@misc{ref52,
  title={Rubidium 85 {D} Line Data},
  author={Steck, D. A.},
  year={2025},
  howpublished={available online at \url{https://steck.us/alkalidata}},
  note={Revision 2.3.4}
}

@article{ref53,
  title={Relativistic many-body calculations of electric-dipole matrix elements, lifetimes, and polarizabilities in rubidium},
  author={Safronova, M. S. and Williams, C. J. and Clark, C. W.},
  journal={Phys. Rev. A},
  volume={69},
  pages={022509},
  year={2004},
  doi={10.1103/PhysRevA.69.022509}
}

@article{ref54,
  title={Using a pair of rectangular coils in the {MOT} for the production of cold atom clouds with large optical density},
  author={Lin, Y.-W. and Chou, H.-C. and Dwivedi, P. P. and Chen, Y.-C. and Yu, I. A.},
  journal={Opt. Express},
  volume={16},
  pages={3753--3761},
  year={2008},
  doi={10.1364/OE.16.003753}
}

@article{ref55,
  title={Gradient echo memory in an ultra-high optical depth cold atomic ensemble},
  author={Sparkes, B. M. and Bernu, J. and Hosseini, M. and Geng, J. and Glorieux, Q. and Altin, P. A. and Lam, P. K. and Robins, N. P. and Buchler, B. C.},
  journal={New J. Phys.},
  volume={15},
  pages={085027},
  year={2013},
  doi={10.1088/1367-2630/15/8/085027}
}

@article{ref56,
  title={Ultra-sensitive measurement of small optical rotation angles using quantum entanglement based on a quasi-{Wollaston} prism beam splitter},
  author={Wang, Shuai and Zhu, Jing and Zhu, Lianqing},
  journal={Opt. Express},
  volume={32},
  number={11},
  pages={19175--19195},
  year={2024},
  doi={10.1364/OE.525608}
}

@article{ref57,
  title={Noise in homodyne and heterodyne detection},
  author={Yuen, H. P. and Chan, V. W. S.},
  journal={Opt. Lett.},
  volume={8},
  pages={177},
  year={1983},
  doi={10.1364/OL.8.000177}
}

@article{LiuDephasing2020,
  title={Experimental demonstration of quantum interference modulation via precise dephasing control in atoms},
  author={Liu, Jinhong and Wu, Jinze and Hu, Yaodong and Zhang, Yu and Zhang, Junxiang},
  journal={Opt. Commun.},
  volume={466},
  pages={125655},
  year={2020},
  doi={10.1016/j.optcom.2020.125655}
}

@article{WrightRaman2008,
  title={{Raman} coupling of {Zeeman} sublevels in an alkali-metal {Bose-Einstein} condensate},
  author={Wright, K. C. and Leslie, L. S. and Bigelow, N. P.},
  journal={Phys. Rev. A},
  volume={78},
  pages={053412},
  year={2008},
  doi={10.1103/PhysRevA.78.053412}
}

@article{LiMicrowaveEIT2009,
  title={Electromagnetically induced transparency controlled by a microwave field},
  author={Li, Hebin and Sautenkov, Vladimir A. and Rostovtsev, Yuri V. and Welch, George R. and Hemmer, Philip R. and Scully, Marlan O.},
  journal={Phys. Rev. A},
  volume={80},
  pages={023820},
  year={2009},
  doi={10.1103/PhysRevA.80.023820}
}

@article{ManjappaDelta2014,
  title={Effects of temperature and ground-state coherence decay on enhancement and amplification in a {$\Delta$} atomic system},
  author={Manjappa, Manukumara and Undurti, Satya Sainadh and Karigowda, Asha and Narayanan, Andal and Sanders, Barry C.},
  journal={Phys. Rev. A},
  volume={90},
  pages={043859},
  year={2014},
  doi={10.1103/PhysRevA.90.043859}
}
\end{document}